\documentclass[lettersize,journal]{IEEEtran}
\IEEEoverridecommandlockouts

\usepackage{cite}
\usepackage[T1]{fontenc}
\usepackage[latin9]{inputenc}
\usepackage{units}
\usepackage{amsmath}
\usepackage{amsthm}
\usepackage{amssymb}
\usepackage{graphicx}
\usepackage{algorithm}  
\usepackage{algorithmic}  
\usepackage{diagbox}
\usepackage[unicode=true,
 bookmarks=false,bookmarksnumbered=true,bookmarksopen=true,bookmarksopenlevel=1,
 breaklinks=false,pdfborder={0 0 0},pdfborderstyle={},backref=false,colorlinks=false]
 {hyperref}
\hypersetup{pdftitle={Your Title}, draft, 
 pdfauthor={Your Name},
 pdfpagelayout=OneColumn, pdfnewwindow=true, pdfstartview=XYZ, plainpages=false}
 \usepackage{hyperref}
 \makeatletter
\g@addto@macro{\UrlBreaks}{\UrlOrds}
\makeatother
\usepackage{flushend}

\usepackage[caption=false,font=footnotesize]{subfig}
\usepackage{float}
\usepackage{subfig}
\usepackage{textcomp}
\usepackage{xcolor} 
\usepackage[printonlyused]{acronym}
\usepackage{multirow}
\newtheorem{theorem}{Theorem}

\providecommand{\lemmaname}{Lemma}
\providecommand{\theoremname}{Theorem}
\usepackage{color}
\usepackage{soul}
\usepackage{marvosym}

\makeatletter

\DeclareRobustCommand*{\lyxarrow}{%
\@ifstar
{\leavevmode\,$\triangleleft$\,\allowbreak}
{\leavevmode\,$\triangleright$\,\allowbreak}}
\providecommand{\tabularnewline}{\\}

\theoremstyle{plain}

\theoremstyle{plain}

\usepackage[caption=false,font=footnotesize]{subfig}

\@ifundefined{showcaptionsetup}{}{%
 \PassOptionsToPackage{caption=false}{subfig}}
\usepackage{subfig}
\makeatother

\providecommand{\lemmaname}{Lemma}
\newtheorem{proposition}{Proposition}
\newtheorem{corollary}{Corollary}

\providecommand{\theoremname}{Theorem}

\def\BibTeX{{\rm B\kern-.05em{\sc i\kern-.025em b}\kern-.08em
    T\kern-.1667em\lower.7ex\hbox{E}\kern-.125emX}}

\begin{document}
\newacro{P2P}{Peer to Peer}
\newacro{PBFT}{Practical Byzantine Fault Tolerance}
\newacro{FBFT}{Fast Byzantine Fault Tolerance}
\newacro{BFT}{Byzantine Fault Tolerance}
\newacro{OCE}{Ordered Cross-shard Endorsement} 
\newacro{SGX}{Software Guard Extensions}
\newacro{PoW}{Proof of Work}
\newacro{PoS}{Proof of Stake}
\newacro{IoT}{Internet of Things}
\newacro{TEE}{Trusted Execution Environment}
\newacro{AWS}{Amazon Web Services} 
\newacro{TC}{Traffic Control}
\newacro{2PL}{two-phase locking}
\newacro{OCC}{optimistic concurrency control}

\title{\huge StableShard: Stable and Scalable Blockchain Sharding with High Concurrency via Collaborative Committees
}

\author{        Mingzhe~Li,~\IEEEmembership{Member,~IEEE}, 
You~Lin, 
and~Jin~Zhang,~\IEEEmembership{Member,~IEEE}
        
\IEEEcompsocitemizethanks{
\IEEEcompsocthanksitem M. Li is with the School of Computing and Information Technology, Great Bay University, Dongguan 523000, China, and with the Institute of High Performance Computing, A*STAR, Singapore (email: mlibn@connect.ust.hk).
\IEEEcompsocthanksitem Y. Lin is with the Department of Computer Science and Engineering, Southern University of Science and Technology, Shenzhen 518055, China (email: liny2021@mail.sustech.edu.cn).
\IEEEcompsocthanksitem J. Zhang is with the Department of Computer Science and Engineering, Southern University of Science and Technology, Shenzhen 518055, China, and also
with Peng Cheng Laboratory, Shenzhen 518055, China (email: zhangj4@sustech.edu.cn).
\IEEEcompsocthanksitem  M. Li and Y. Lin are the co-first author.
\IEEEcompsocthanksitem J. Zhang is the corresponding author.
}
}

\markboth{IEEE/ACM Transactions on Networking, VOL. XX, NO. XX, XX 2026}%
{Lin \MakeLowercase{\textit{et al.}}: StableShard: Stable and Scalable Blockchain Sharding with High Concurrency via Collaborative Committees}
\IEEEtitleabstractindextext{

\begin{abstract}

Sharding enhances blockchain scalability by partitioning nodes into multiple groups for concurrent transaction processing.
Configuring a large number of \emph{small shards} usually helps improve transaction concurrency, but it also increases the fraction of malicious nodes in each shard, easily causing shard corruption and jeopardizing system security.
Existing works attempt to improve concurrency by reducing shard sizes while maintaining security, but typically rely on \emph{time-consuming recovery of corrupted shards to restore liveness} and \emph{network-wide consensus}. This causes severe system stagnation and limits scalability.

To address this, we present StableShard, a sharded blockchain that securely provides high concurrency with stable and scalable performance.
The core idea is to carefully co-design the division of labor between proposer shards (PSs) and finalizer committees (FCs): we \emph{deliberately assign 1) asymmetric roles and 2) matching parameters} to PSs and FCs.
Small PSs focus on fast transaction proposal and local validity, while large FCs focus on resolving forks, finalizing PS blocks, and maintaining liveness for faulty PSs via a cross-layer view-change protocol.
Moreover, by fine-tuning key system parameters (e.g., shard size, quorum size), we ensure each PS to tolerate <1/2 fraction of malicious nodes without lossing liveness, and allow multiple FCs to securely coexist (each with <1/3 fraction of malicious nodes) for better scalability.
Consequently, StableShard can safely configure many smaller PSs to boost concurrency, while FCs and PSs jointly guarantee safety and liveness \emph{without system stagnation}, leading to stable and scalable performance.
Evaluations show that StableShard achieves up to $10{\times}$ higher throughput than existing solutions and significantly more stable concurrency under attacks.

\end{abstract}

}

\maketitle

\IEEEdisplaynontitleabstractindextext

\IEEEpeerreviewmaketitle

\section{Introduction}\label{sec:introduction}

\IEEEPARstart{B}{lockchain} 
sharding has attracted widespread attention as a technique to address low scalability in traditional blockchain~\cite{bitcoin, eth}.
Its main idea is to partition the blockchain network into smaller groups, known as shards~\cite{hellings2021byshard, cheng2024shardag, tao2020sharding, tao2023sharding, zheng2021meepo, amiri2021sharper, ruan2021blockchains, omniledger, rapidchain, pyramid, monoxide, li2022jenga, cycledger, repchain, sgxSharding}. 
Each shard manages a unique subset of the blockchain ledger state and performs intra-shard consensus to produce blocks concurrently.
Generally, \emph{configuring a larger number of smaller shards tends to result in better transaction concurrency} under the same network size~\cite{gearbox, instachain, cochain}.

However, existing permissionless sharding systems require \emph{large shard sizes} to ensure security, significantly limiting transaction concurrency in large-scale blockchain sharding systems~\cite{omniledger, Harmony, zilliqa}. 
In most permissionless blockchain sharding systems, nodes are \emph{randomly assigned} to disjoint shards~\cite{omniledger, rapidchain, repchain, pyramid, li2022jenga, sschain}. 
This randomness causes \emph{smaller shards more likely to contain a larger fraction of malicious nodes} that exceed the fault tolerance threshold (e.g., $ \geq 1/3$ for BFT-typed intra-shard consensus mechanism), resulting in shard corruption~\cite{beyondOneThird} and compromised system security. 
Consequently, current systems tend to configure large shard sizes (e.g., 600 nodes per shard in OmniLedger~\cite{omniledger}) to substantially limit the probability of each shard's corruption~\cite{omniledger, zilliqa, Harmony}. 
Unfortunately, such large shard sizes not only slow down intra-shard consensus but also decrease the network's overall shard count, leading to reduced system transaction concurrency.

While some previous studies have attempted to reduce shard size to improve concurrency, their solutions have various limitations. 
Some works make less practical assumptions (e.g., synchronous network \cite{rapidchain, repchain, cycledger, li2020polyshard, li2025sp}, reduced resiliency \cite{pyramid, li2022jenga}, small scale \cite{amiri2021sharper, hellings2021byshard, das2020efficient, zheng2021meepo, tao2023sharding, tao2020sharding, cheng2024shardag}) or require specialized hardware~\cite{sgxSharding, SGX}, limiting practicality and deployability.
More closely related, recent designs reduce shard sizes by tolerating corrupted shards (i.e., shards with a larger fraction of malicious nodes). 
However, they typically recover by triggering explicit recovery events after liveness breaks (e.g., replacing a faulty shard \cite{cochain} or resampling a shard with larger parameters after detecting liveness failure \cite{gearbox}). 
Unfortunately, the systems lose liveness during recovery, leading to severe \emph{\textbf{temporary stagnation}} issues in real-world scenarios, and they are fundamentally different from keeping every small shard continuously live \emph{in-place}.
For example, our measurements in Section~\ref{subsec:shard_size_security} show that with 24 shards, migrating pruned Ethereum \cite{chainDataETH} historical states can take over \emph{100 minutes} per node.
Moreover, some of these works~\cite{rana2022free2shard, gearbox} rely on network-wide consensus to ensure security, \emph{\textbf{limiting scalability}}.

\begin{figure}[t]
\centerline{\includegraphics[width=0.39\textwidth]{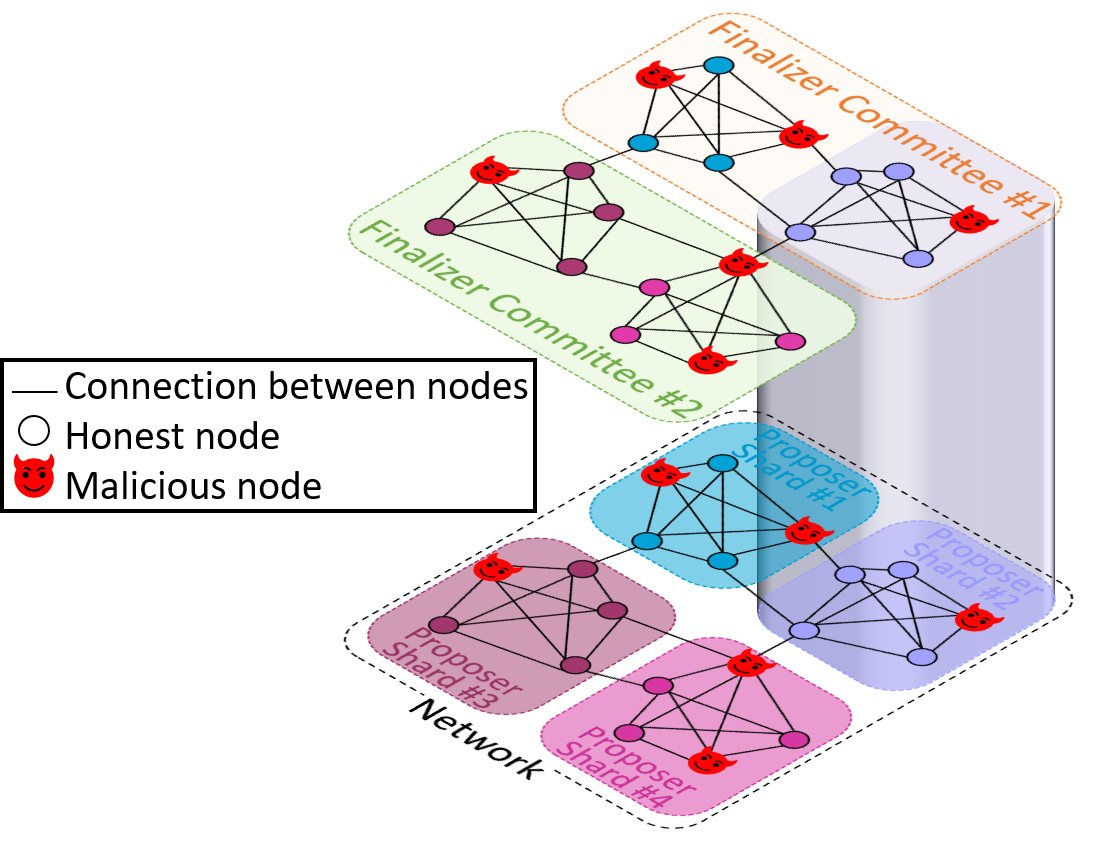}}
\caption{Illustration of the Division-of-Labor Collaboration. A node simultaneously belongs to a PS and a FC.}
\label{fig:overview}
\end{figure}

\vspace{6pt}
This paper proposes StableShard, a highly concurrent blockchain sharding system that fills the above gap: it provides \emph{stable and scalable performance} under corrupted shards \emph{without} network-wide consensus and \emph{without} any recovery procedure that leads to system stagnation.
Our core idea is a Division-of-Labor Collaboration (DoLC) protocol, as shown in Figure~\ref{fig:overview}. 
We \emph{co-design how proposer shards (PSs) and finalizer committees (FCs) divide responsibilities and collaborate} so that safety and liveness are guaranteed by complementary mechanisms.
Concretely, we deliberately assign \emph{\textbf{asymmetric roles}} and \emph{\textbf{matching parameters}} to PSs and FCs.
Many small PSs are optimized for concurrency: they propose blocks quickly and guarantee transaction validity through a carefully chosen intra-shard quorum.
Multiple larger FCs are optimized for robustness and scalability: each FC finalizes multiple disjoint PSs by running lightweight BFT on headers to resolve forks and provide finality, avoiding network-wide consensus.
Crucially, FCs also maintain \emph{stagnation-free liveness} for faulty PSs via a cross-layer view-change protocol, so that even corrupted PSs remain live \emph{in-place} instead of triggering recovery events.
Moreover, by fine-tuning key parameters (e.g., committee sizes, quorum thresholds) jointly, StableShard enables PSs to tolerate a Byzantine fraction approaching $<1/2$ while keeping each FC secure ($<1/3$) with overwhelmingly high probability, thereby supporting many small PSs and multiple coexisting FCs for stable and scalable concurrency.

\vspace{6pt}
Realizing the above goal requires addressing the following challenges.

\vspace{6pt}
\noindent
\textbf{Challenge 1: Scalable Finalization without Network-wide Consensus.} 
The first challenge is how to ensure security despite the presence of corrupted shards while \emph{\textbf{preserving scalability}}.
To tackle this challenge, we propose a \emph{scalable PS-FC Division-of-Labor Collaboration (DoLC) protocol}:
Numerous smaller PSs process transactions and propose blocks through intra-shard consensus.
However, their small size makes them more prone to corruption. 
Those corrupted PSs may fail to reach consensus (e.g., create forks) and jeopardize system security.
Therefore,  multiple larger FCs are set to verify and perform BFT-typed consensus on the consensus results of the corresponding PSs to finalize their transactions and eliminate forks (via our \emph{fork selection rules}).
To balance scalability and security, we \emph{prudently fine-tune FC sizes} to the minimum with negligible failure probability.
This implies multiple FCs co-exist, and the proportion of malicious nodes in each FC is less than $1/3$ (i.e., this FC is secure and honest) with a extremely high probability. 
As a result, each honest FC can provide finality for multiple PSs, mitigating network-wide consensus and enhancing scalability.

\vspace{6pt}
However, it is difficult to achieve stable performance and efficiency if the system only relies on the FCs to ensure security (i.e., both liveness and safety). 
This is where the limitations of some existing works come into play \cite{sschain, pyramid, gearbox, cochain, rana2022free2shard}.
To address this challenge, we differentiate ourselves by requiring FCs and PSs to \emph{divide responsibilities and collaborate to guarantee liveness and safety, thus providing stable performance and reducing overhead}.
Detailed explanations are as follows.

\vspace{6pt}
\noindent
\textbf{Challenge 2: Stagnation-free Liveness without Recovery for Stable Performance.} 
The second challenge is how to maintain \emph{\textbf{stable performance}}.
In simplistic approaches~\cite{gearbox, cochain}, corrupted shards that lose liveness require replacement or reshuffling. 
This results in time-consuming cross-shard state migration~\cite{sschain} and significant system stagnation issues.
To tackle this challenge, we \emph{ensure that every PS (even when it is corrupted) will not lose liveness with FC's help and quorum co-design}.
We identify two scenarios wherein a corrupted shard may lose liveness.
Firstly, a malicious leader may cause a PS to lose liveness by not proposing blocks.
In such instances, a corrupted PS cannot rely on itself to perform the view change \cite{pbft} to replace the malicious leader.
To address this, we propose a \emph{cross-layer view change protocol} that uses FCs to replace the malicious leaders for PSs.
Secondly, when the number of honest nodes in a shard falls below the quorum size (the number of votes required for reaching consensus), malicious nodes can cause the shard to lose liveness by remaining silent.
Consequently, we \emph{deliberately configure} 
\begin{equation}
\label{intro_expression_1}
\# \ of \   honest \ nodes \geq quorum \ size
\end{equation}
within each PS, achieved through rigorous theoretical derivations, to guarantee the consensus process for valid blocks can be sustained.
These designs \emph{\textbf{ensure liveness}} within each PS, \emph{\textbf{avoiding recovery}} processes, and preventing system performance degradation due to stagnation.

\vspace{6pt}
\noindent
\textbf{Challenge 3: Low-overhead Safety via Complementary Responsibilities.} 
The third challenge is how to achieve \emph{\textbf{low overhead}} for cross-layer communication in DoLC while ensuring safety.
When we rely exclusively on FCs for safety, FCs must obtain entire blocks from PSs to verify raw transactions and solve the forking issue, leading to considerable overhead.
To address this challenge, our key idea is to \emph{rely on the consensus within PSs to guarantee the validity of raw transactions and utilize FCs to resolve the forking issue of PSs, so that \textbf{only headers} are transmitted across layers, reducing overhead.} 
We observe that if the number of malicious nodes is less than the quorum size, a block containing invalid transactions will not pass the intra-shard consensus, as it cannot collect enough votes (honest nodes do not vote for invalid blocks). 
Therefore, our solution is to guarantee 
\begin{equation}
\label{intro_expression_2}
quorum \ size > \# \ of \  malicious \ nodes
\end{equation}
in each PS (guarantee transaction validity), guided by \emph{careful parameterization} and rigorous theoretical calculations.
However, a corrupted PS can still fork its chain in this case. 
Luckily, such a safety attack can be detected by checking block headers. 
Therefore, we require FCs to \emph{\textbf{obtain block headers}} from the corresponding PSs, select one fork branch for each PS, and finalize it through consensus. 
These designs \emph{\textbf{ensure safety}} with \emph{\textbf{low overhead}}.

\vspace{6pt}
\noindent
\textbf{Contributions.} The contributions of this work are as follows: 
\begin{itemize}
  \item 
  We present StableShard, a sharded blockchain that achieves \emph{high concurrency with stable and scalable performance}. StableShard eliminates recovery procedures that cause stagnation and avoids network-wide consensus.
  
  \item 
  We co-design a scalable PS-FC Division-of-Labor Collaboration (DoLC) protocol. Small PSs maximize concurrency and guarantee transaction validity via quorum voting. To provide stable performance, FCs maintain in-place liveness for faulty PSs via a cross-layer view-change mechanism. For balanced security and scalability, multiple large and secure FCs provide header-only fork resolution and finality for multiple PSs. 


  \item
  We introduce a joint parameterization (committee sizes and quorum thresholds). By jointly satisfying expressions~\ref{intro_expression_1} and ~\ref{intro_expression_2}, and together with the DoLC protocol, StableShard tolerates $< 1/2$ (instead of $< 1/3$) fraction of malicious nodes per PS while guaranteeing \emph{stagnation-free liveness}. It also enables multiple secure FCs ($<1/3$ fraction of malicious nodes) to coexist. Jointly, StableShard achieves high concurrency, stable and scalable performance.
  

  \item We implement StableShard based on Harmony~\cite{Harmony}, an industrial sharding codebase once had a top 50 market cap in cryptocurrency, and evaluate it at scale (up to 2,550 nodes). Results show up to $10\times$ throughput improvement and significantly more stable concurrency under attacks over the baseline system GearBox (CCS 22) \cite{gearbox}.

\end{itemize}

\section{Background and Related Work}

\subsection{Background on Blockchain Sharding}
\label{subsec:blockchain_sharding}

Sharding has been extensively studied as a promising solution for improving the scalability of blockchain~\cite{hellings2021byshard, cheng2024shardag, tao2020sharding, tao2023sharding, zheng2021meepo, amiri2021sharper, ruan2021blockchains, omniledger, rapidchain, pyramid, monoxide, li2022jenga, cycledger, repchain, sgxSharding}.
Its core idea is partitioning nodes into groups (aka shards). 
Each shard maintains a disjoint subset of the states and reaches intra-shard consensus to process disjoint transactions in parallel. 
A blockchain sharding system usually has the following main components.

1) \emph{Shard Formation between Epochs}: A blockchain sharding system typically operates in fixed periods named \emph{epochs} (e.g., one day).
Initially, the system imposes some restrictions (e.g., 
Proof of Stake) to decide the nodes that join the network to prevent Sybil attacks.
Once epoch participants are identified, the system typically assigns nodes across shards using public-variable, bias-resistant, and unpredictable \emph{epoch randomness}. 
This prevents malicious nodes from grouping into a single shard.

2) \emph{Intra-shard Consensus}: After shard formation, each shard makes intra-shard consensus to append blocks into its shard chain.
Most systems~\cite{cheng2024shardag, zheng2021meepo, rapidchain, pyramid, li2022jenga, gearbox} adopt BFT-typed consensus protocol (e.g.,  Practical Byzantine Fault Tolerance, PBFT) to produce blocks~\cite{sok}.
In such consensus protocols, honest nodes vote for valid blocks and will only accept blocks for which quorum size of votes have been collected.

\emph{StableShard applies BFT-typed intra-shard consensus protocol, as many other sharding systems do.}
The BFT-typed intra-shard consensus protocol is a pluggable component.
For implementation simplicity and fair comparison, we use the \ac{FBFT} consensus protocol proposed by Harmony~\cite{Harmony} in this paper.
FBFT is a variation of PBFT~\cite{pbft}, a leader-based consensus protocol providing the same security guarantee as PBFT.
It, by default, withstands $<m/3$ malicious nodes with $2m/3 + 1$ quorum size in a group of size $m$ under a partial-synchronized network.
Moreover, like existing BFT-type consensus protocols in partially synchronous networks, FBFT also incorporates a \emph{timeout} and \emph{view change} mechanism to ensure consensus liveness. Similarly, during the asynchronous phase before GST, the timeout duration gradually increases with each view change and eventually converges to a stable value (once the network becomes synchronous), to accommodate the asynchronous nature of the network and prevent overly frequent view changes~\cite{pbft, hotstuff, pyramid}.



3) \emph{Cross-shard Mechanism}: 
Sharding partitions the ledger among shards, necessitating cross-shard transaction mechanisms to update each shard atomically. 
StableShard applies the relay-based mechanism similar to Monoxide~\cite{monoxide} and Harmony~\cite{Harmony}, 
which initially packages cross-shard transactions in the source shard to deduct and forwards them to the destination shard with deduction proof. 
Then, the destination shard does the deposit operations. 
This protocol provides eventual atomicity and asynchronously lock-free processing to cross-shard transactions, preserving scalability even if almost all transactions are cross-shards~\cite{monoxide}.


\subsection{Blockchain Sharding for Improved Concurrency}
\label{subsec:shard_size_security}

We group the most relevant prior work into three categories according to how they trade concurrency, safety, and liveness, and how they handle corrupted shards and performance stability.

1) \emph{Permissionless sharding with conservative (large) shards.}
Classic permissionless sharding systems (e.g., \cite{elastico, omniledger, Harmony}) improve concurrency by parallelizing intra-shard consensus, but typically require large shard sizes so that each shard remains honest with overwhelming probability.
This conservative choice preserves safety/liveness under the standard $<1/3$ BFT threshold, yet it \emph{limits scalability because large shards slow consensus and reduce the number of shards}, thus capping concurrency.

2) \emph{Small-shard sharding under stronger assumptions or weakened resilience.}
A separate line of work reduces shard sizes via additional assumptions or auxiliary support, e.g., assuming synchrony within shards~\cite{rapidchain, repchain, cycledger, li2020polyshard, li2025sp} or leveraging trusted hardware~\cite{sgxSharding}.
Some works~\cite{pyramid, li2022jenga} achieve smaller shard sizes at the cost of overall system resilience.
Some other designs achieve very small shards by restricting the overall network scale or not rigorously accounting for corruption probability at large scale \cite{amiri2021sharper, hellings2021byshard, das2020efficient, zheng2021meepo, tao2023sharding, tao2020sharding, cheng2024shardag}.
These approaches can boost concurrency, but \emph{their assumptions (synchrony, extra hardware, weakened resilience, or small deployment scale) reduce practicality} in large permissionless settings.

3) \emph{Corrupted-shard-tolerant designs and hierarchical/decoupled confirmations.}
Only a few blockchain sharding systems explicitly permit corrupted shards to shrink committees while attempting to preserve overall security.
Free2Shard~\cite{rana2022free2shard} relies on network-wide coordination (and PoW-style assumptions), which \emph{limits deterministic finality and can constrain scalability}.
GearBox~\cite{gearbox} leverages the safety-liveness dichotomy and detects liveness violations to trigger "gear switching" (i.e., adjusting committee parameters, injecting more nodes) so that liveness can be restored.
However, such \emph{recovery is orchestrated at a network-wide level and can introduce instability} under repeated or adaptive attacks, as shards may frequently enter the recovery path. 
Similarly, systems that rely on network-wide coordination \cite{sschain, rana2022free2shard} tend to face scalability limits due to global monitoring/consensus overhead.
CoChain~\cite{cochain} monitors each shard using groups of other shards and recovers the system via shard replacement/takeover once a corrupted shard is found (Consensus-on-Consensus). 
While effective, this recovery model is fundamentally different from keeping small shards continuously live in-place, and it still \emph{raises practical costs for transferring historical ledgers/transaction context to the replacement executor.}

\vspace{6pt}
\noindent
\textbf{Severe Stagnation from Recovery/State Movement.}
A common Achilles' heel of many corrupted-shard-tolerant designs is that liveness recovery is coupled with heavy recovery actions (e.g., state/ledger transfer, reshuffling, executor replacement).
Even when safety can be maintained, these recovery paths may pause service and cause large latency spikes, which is unacceptable for stable performance.

\begin{figure}[t]
\centerline{\includegraphics[scale=0.33]{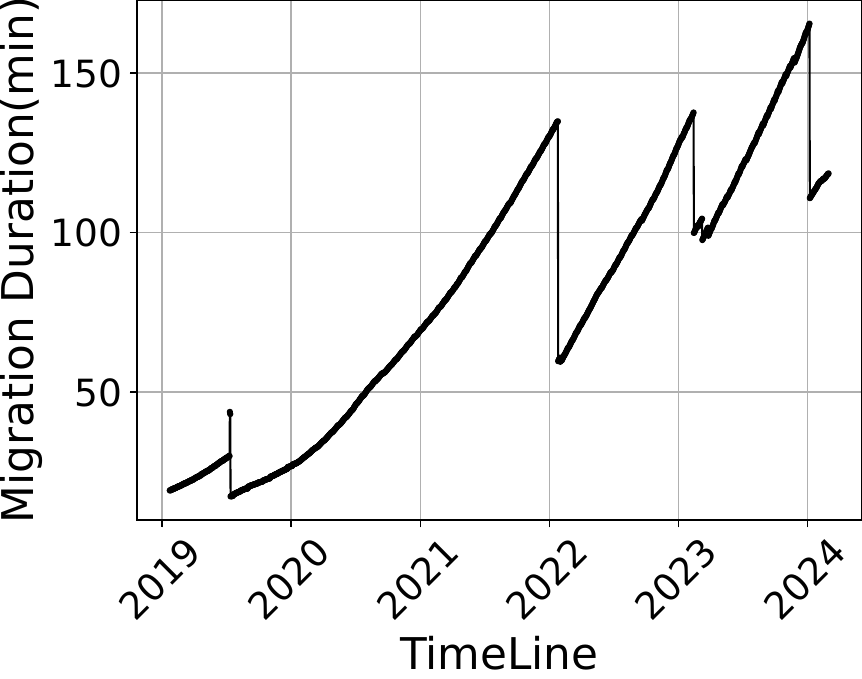}}
 \caption{Migration duration of Ethereum historical ledgers.} 
\label{fig:eth_migration}
\end{figure}

We conducted estimations of substantial costs associated with state migrations. 
Based on the data recorded on the Etherscan, the size of the archive chain data has exceeded 16TB.
Even if the node is in pruning mode to synchronize the data, the data volume is over 970GB.
We conducted simulations to estimate the time required for a node with a bandwidth of 50Mbps to execute the state migration of Ethereum~\cite{eth}. 
The findings reveal that, notwithstanding ledger pruning and utilizing 24 shards, the migration process has extended beyond 100 minutes since 2024, as shown in Figure~\ref{fig:eth_migration}. 

Furthermore, estimating the migration procedure, even with the download limited to the state trie for verifying new transactions, consumes more than 25 minutes~\cite{stateTrie, headState}.
Such stagnation issues significantly hamper system functionality, causing a notable decline in performance and user experience.

\vspace{6pt}
\noindent
\textbf{Our Positioning.}
In StableShard, we also permit corrupted proposer shards, but we differ from the above lines in a key way: we \emph{avoid recovery-after-failure} altogether.
Instead of respawning/resizing shards after liveness breaks (GearBox), or replacing/taking over corrupted shards after detection (CoChain), StableShard keeps shard membership and states intact and maintains progress \emph{\textbf{in-place}} through PS-FC cooperation.

Concretely, StableShard co-designs (i) a complementary division of labor and (ii) matched parameters:
PSs guarantee transaction validity locally via quorum voting, enabling FCs to remain stateless and finalize using only headers for fork-resolution and finality;
FCs additionally maintain \emph{stagnation-free liveness} for faulty PSs via a cross-layer view-change protocol, so a corrupted PS does not trigger reshuffling or state migration.
By jointly fine-tuning committee sizes and quorum thresholds, StableShard supports many small PSs (tolerating Byzantine fraction approaching $<1/2$) while allowing multiple FCs to coexist securely (each $<1/3$), thus achieving high concurrency with stable and scalable performance.

\section{The StableShard Model} 
\subsection{Network Model}
\label{section:network}
StableShard is deployed on a \emph{partial-synchronous} Peer to Peer network, in which there is a known bound $\delta$ and an unknown Global Stabilization Time (GST).
After GST, the network becomes synchronous, and all transmissions between two honest nodes arrive within time $\delta$~\cite{hotstuff}.
As is common with most previous systems~\cite{bitcoin, Harmony, li2022jenga}, messages in StableShard are propagated via a gossip protocol.

StableShard consists of $N$ nodes, each having a public/private key pair representing its identity when sending messages.
Each node belongs to one proposer shard (PS) in the second layer and one finalizer committee (FC) in the first layer simultaneously.
There are $C$ FCs, each composed of $K$ disjoint PSs.
Hence, the system comprises $C \cdot K$ PSs.
Besides, each FC has $n=N/C$ nodes, and each PS has $m=n/K$ nodes.

Each \emph{proposer shard}, similar to shards in traditional sharding, is responsible for transaction processing. 
It runs BFT-typed intra-shard consensus to append proposer blocks recording transactions to its proposer chain.
Each \emph{finalizer committee} comprises multiple PSs from the second layer and verifies their proposer blocks.
It runs BFT-typed consensus to append finalizer blocks recording hashes of proposer blocks (but not raw transactions) to its finalizer chain.
The consensus running in each FC is lightweight, as it does not have to handle heavy transactions.

\subsection{Transaction Model}
StableShard adopts the account/balance model to present the ledger state, the same as existing works~\cite{eth, pyramid, monoxide}. 
The state (i.e., balance) of a given account is maintained by a single PS in StableShard based on the hash of the account address.
Each transaction is routed to the corresponding PS based on the related account addresses for processing.
Without loss of generality, we will illustrate our system based on normal transfer transactions.
However, StableShard is \emph{also compatible with handling smart contracts}, as discussed in Section~\ref{subsec:additional_discussion}.

\subsection{Threat Model}
\label{section:threat_model}
In StableShard, two categories of nodes exist: honest and malicious. 
Honest nodes comply with all StableShard protocols. 
For example, they actively respond to consensus messages, refuse to sign blocks containing invalid transactions, and consistently broadcast their signed messages to the entire network.
On the other hand, malicious nodes can conduct various types of attacks.
In StableShard, three main typical attacks can have an additional impact on our system security:
(1) silent attack, where they refuse to respond to messages to disrupt consensus;
(2) transaction manipulation attack, where they can include invalid transactions into blocks; 
(3) equivocation attack, where they can send different messages to different nodes.
For other typical attacks (e.g., transaction censorship~\cite{censorship}, eclipse~\cite{eclipse}, etc.), there are solutions~\cite{redBelly, honeybadger} that are orthogonal to our main designs and can be adopted by StableShard (refer to Section~\ref{subsec:additional_discussion} for more discussion).
Besides, it is assumed that malicious nodes cannot forge messages by accessing other nodes' private keys. 
We denote the fraction of malicious nodes in the system as $f$, indicating $f \cdot N$ malicious nodes. 
Similar to existing sharding systems~\cite{pyramid, li2022jenga, cochain}, we operate under the assumption of slowly-adaptive adversaries that the distribution of honest and malicious nodes remains static within each epoch (typically one day), and alterations can only occur between epochs.
\section{Division-of-Labor Collaboration Protocol} 

The PS-FC Division-of-Labor Collaboration (DoLC) protocol is the backbone component that ensures security while allowing smaller shards for improved concurrency without sacrificing scalability.
Specifically, it creates numerous smaller proposer shards for high transaction concurrency and establishes larger finalizer committees to safeguard potentially corrupted PSs.
As the number of nodes increases, our system does not rely on network-wide consensus but rather safely establishes additional FCs, each responsible for the security of a subset of PSs, thereby ensuring scalability.

This section describes the \emph{\textbf{basic design}} of our DoLC protocol that accomplishes the above goals.
However, the system still faces problems on top of this basic architecture. 
First, the system faces the problem of loss of liveness, which leads to system stagnation and inability to provide stable performance.
Second, relying solely on FCs to safeguard the safety of PSs would introduce considerable overhead. 
Third, since our system may encounter corrupt proposer shards, handling cross-shard transactions poses a unique challenge. 
Finally, a well-developed system should further guarantee efficiency.
We will describe how we deal with these difficult challenges in Section~\ref{sec:design}.

\subsection{Formation of PSs and FCs}
\label{sec:formation_of_architecture}
StableShard runs in fixed periods called \emph{epochs} with a duration according to the system requirements (e.g., one day for most existing blockchain sharing systems~\cite{pyramid, rapidchain, cochain}).
StableShard applies \ac{PoS} that requires nodes to stake a certain amount of tokens to join the epoch to prevent Sybil attacks, similar to existing works~\cite{Harmony, li2022jenga}.
We assume a trusted beacon chain publically records the identities of these nodes for each epoch, similar to most blockchain sharding~\cite{pyramid, omniledger, li2022jenga}.
StableShard utilizes epoch randomness to assign nodes to PSs randomly to prevent malicious nodes from gathering.
The generation of epoch randomness leverages a combination of verifiable random function (VRF)~\cite{micali1999verifiable} and verifiable delay function (VDF)~\cite{boneh2018verifiable} techniques, as outlined in prior work~\cite{Harmony}.

Unlike most existing sharding systems, StableShard is designed as a division-of-labor and collaboration between PSs and FCs, with each node belonging to a PS and a FC simultaneously.
When given a network size $N$ and a fraction of malicious node $f$, the system configures FC size $n$ and PS size $m$, ensuring a negligible probability of failure to obtain the number of PSs securely.
Subsequently, the FC identifier of a node is obtained based on the PS identifier of the node (i.e., without loss of generality, a node in $PS_j$ belongs to $FC_{\lfloor(j+K-1)/K\rfloor }$ where $K$ represents the number of PSs per FC).
In other words, $FC_{i}$ is responsible to safeguard $PS_j$ where $j \in [(i-1)K+1,iK]$.

\subsection{Strawman Design of DoLC protocol}
\label{sec:layered_consensus}
We now introduce the strawman design of StableShard's DoLC protocol. 
The DoLC protocol will be more efficient when incorporating the various component designs in Section~\ref{sec:design}.
The basic DoLC protocol comprises the following three phases:

\vspace{6pt}
\noindent
\emph{Block Proposal.}
In this phase, each PS executes intra-shard consensus to append proposer blocks containing transactions to its proposer chain. 
As Section~\ref{subsec:blockchain_sharding} mentions, we use FBFT~\cite{Harmony} as the intra-shard consensus.
However, we derive a different quorum size $m/2 + 1 $ for a PS of size $m$ (the rationale is shown in Section~\ref{subsec:overhead}. PSs cannot guarantee safety on their own, that is why they need FCs' assistance).
Like most systems, honest nodes verify raw transactions 
before voting for proposer blocks.
Nevertheless, a PS may be corrupted due to its smaller size. 
Hence, each honest node must broadcast the proposer block (\emph{after the designs in Section \ref{subsec:overhead}, only header is required}) it voted for to the corresponding FC for verification and finalization later.
This design guarantees the block will be broadcast to the FC if it passes the intra-shard consensus and is voted by at least one honest node.

\vspace{6pt}
\noindent
\emph{Cross-layer Verification.}
During this phase, FCs verify the proposer blocks they receive.
Before diving into the verification process, FC nodes must first \emph{exclude} any proposer blocks that conflict with already finalized proposer blocks. 
This design reduces verification and storage overhead for FC nodes as the FC cannot simultaneously finalize conflicting proposer blocks.
A conflict occurs when the received proposer block is not a successor of the most recently finalized proposer blocks within the same PS.
To achieve this exclusion, FC nodes check the \texttt{parent block} of the received proposer block to check the topological relationship between proposer blocks.

Next, nodes within the FC verify the validity of transactions in proposer blocks. 
Our basic design presumes FC nodes keep track of all corresponding PS states to verify each raw transaction.
\emph{A more efficient design refining this process is detailed in Section~\ref{subsec:overhead}.}
After validation, each FC node retains valid proposer blocks in its memory.

\vspace{6pt}
\noindent
\emph{Block Finalization and Fork Selection.}
In this phase, each FC performs standard BFT-typed consensus (e.g., FBFT, as mentioned in Section~\ref{subsec:blockchain_sharding}, tolerating $< n/3$ malicious nodes with $2n/3 + 1$ quorum size) to produce finalizer blocks containing hashes of valid proposer blocks.
Honest FC nodes verify the validity of proposer blocks recorded in finalizer blocks. 
They can use cached verification results from the previous phase to speed up. 
Additionally, each FC assists its corresponding PSs in fork selection through intra-shard consensus, as will be explained later.
After passing the FC's consensus, honest nodes broadcast a finalizer block to the corresponding PSs.
This allows PSs to confirm the execution of finalized proposer blocks and update the ledger state.

The process by which FCs assist PSs in \textbf{\emph{fork selection}} is critical, as it determines which fork from a potentially corrupted PS will ultimately be finalized by the FC. In partial-synchronous networks, deterministically detecting forks is impossible. Fortunately, StableShard does not require deterministic fork detection. 
Within any FC, the leader follows two primary rules to select a fork from its corresponding PSs. First, the selected fork must be a continuation of the last fork finalized by the FC to prevent conflicts. Second, there may be multiple forks that all extend from the same fork previously finalized by the FC. Among these, the leader selects the longest fork from its own perspective and proceeds to finalize the latest block on this fork, along with other transactions, through intra-shard consensus.
Honest nodes within the FC, when participating in consensus, will verify whether the fork selected by the leader is indeed a continuation of the last fork finalized by the FC and whether the leader has selected only one fork (to prevent equivocation attacks). However, these honest nodes do not judge whether the leader has selected the longest fork, as different nodes may have inconsistent views of the network state in a partial-synchronous environment.
This fork selection mechanism relies on the intra-shard consensus of the FC. Therefore, when the intra-shard consensus within the FC is secure, this mechanism ensures the security of fork selection.

\vspace{6pt}
\noindent
\textbf{Balance of Scalability and Security.}
The DoLC protocol relies on FCs for security. 
We ensure that FC sizes are minimal, preserving an extremely high likelihood of being secure to balance security and scalability.
Section~\ref{sec:epoch_security} details the complete derivation and proof.



\section{Enhanced Designs for StableShard}
\label{sec:design}

Based on the basic DoLC protocol, in this section, we introduce the core and unique designs of StableShard to achieve \emph{stable performance, low overhead, secure cross-shard transaction handling, and high efficiency}.
First, to address the system stagnation issue and provide stable performance, we design a cross-layer view change mechanism and ensure the quorum size within PSs (even if corrupted) can be reached so that each PS will not lose liveness. 
This is introduced in Section \ref{subsec:stagnation}.
Second, to reduce cross-layer communication overhead while ensuring safety, we rely on consensus within PSs to ensure the validity of transactions and then utilize FCs to resolve the forking issues of PSs.
Therefore, only headers are transmitted across two layers for low overhead. 
This part will be explained in Section \ref{subsec:overhead}.
Third, as will be described in Section \ref{subsec:cross-shardconfir}, to guarantee cross-shard transaction security, we require the finalization of cross-shard transactions to depend on secure FCs.
Finally, to achieve higher system efficiency, we design a pipeline mechanism in which PSs and FCs produce blocks concurrently, which will be discussed in Section \ref{subsec:pipeline}.


\subsection{Stagnation Prevention}
\label{subsec:stagnation}
We focus on performance stability in this part. 
The performance may suffer when a corrupted PS mounts a silent attack on liveness. 
The conventional solution of reshuffling non-liveness shards can cause significant delays due to state migration, thus degrading performance. 
Hence, \emph{we aim to guarantee liveness in each PS}, even if corrupted.
We have identified two scenarios where a corrupted PS can successfully execute a silent attack against liveness.

In the \emph{first scenario}, a PS's malicious leader stops block production. 
In most BFT-typed consensus protocols \cite{pbft, hotstuff, Harmony}, leaders package transactions into blocks to be voted for. 
Without block production, the consensus process cannot proceed.
Although the consensus typically has a view change mechanism to substitute a malicious leader, 
a corrupted PS has too many malicious nodes (exceeding the BFT-typed consensus's tolerance threshold, $1/3$) to replace a malicious leader independently.
Hence, we propose the \textbf{\emph{cross-layer view change mechanism}} using FCs to replace the malicious PS leader.
The mechanism includes these steps:


\vspace{6pt}
\noindent 
\emph{Complaint for the Leader.}
Suppose a node has not accepted blocks from the leader within a \emph{timeout} (mentioned in Section \ref{subsec:blockchain_sharding}).
In that case, it suspects leaders of remaining silent and initiates the cross-layer view change by sending \texttt{Complain} message to its PS and its FC.
This message should contain the suspected leader's identifier and the reasons, such as a lack of block proposal.

\vspace{6pt}
\noindent 
\emph{Consensus on Complaint.}
Upon receiving \texttt{Complain} messages from PS nodes, FC nodes verify whether this PS leader's block has not been received.
The FC randomly selects one complainer as the PS' new leader based on epoch randomness if the number of valid \texttt{Complain} messages reach the PS's quorum size (we ensure this, as will be explained in the second scenario).
Through the FC's consensus, the \texttt{Complain} messages and the new leader's identity are packaged into a finalizer block and broadcast to the PSs.

\vspace{6pt}
\noindent 
\emph{Transition of the Leader.}
Upon receiving the consensus results from the FC, the PS nodes update their local view of the leader and follow the new leader for block proposals and intra-shard consensus.

\vspace{6pt}
\noindent
\textbf{Discussion of Cross-Layer View Change.}
The security of this mechanism depends on the FCs, which is the same as our DoLC protocol.
Fortunately, we guarantee the security of FCs (each with $1<3$ fraction of malicious nodes) with high probability after rigorous theoretical derivations and proof, as shown in Section~\ref{sec:epoch_security}.
Besides, this view change mechanism can also replace malicious leaders that launch other attacks (e.g., equivocation and transaction manipulation attacks).
Specifically, honest nodes must provide two blocks of the same height signed by the same leader to prove an equivocation attack or a block containing an incorrect transaction to prove a transaction manipulation attack (after the designs in next Section \ref{subsec:overhead}, only block headers are required to detect equivocation, and the transaction validity can be guaranteed inside each PS).
FCs then can verify the feasibility of \texttt{Complain} messages.

The above mechanism is secure both before and after GST, provided that the intra-shard consensus within the FC is secure. After GST, its security is unequivocal because the network transitions to a synchronous state, and timeouts stabilize.
Before GST, the mechanism remains secure because it relies on the assistance of the secure FC to perform cross-layer view changes for its corresponding PSs. The intra-shard consensus within each FC is no different from existing BFT-type consensus designed for partial-synchronous networks, which are equipped with dynamically increasing timeout mechanisms and robust view change protocols (as detailed in Section \ref{subsec:blockchain_sharding}).
Therefore, even before GST, when the FC is secure (i.e., the proportion of malicious nodes <1/3), its intra-shard consensus can guarantee both safety and liveness (liveness may be temporarily affected due to asynchrony). As a result, each FC can securely assist PSs in cross-layer view change, even before GST.

\vspace{6pt}
In the \emph{second scenario}, the PS has fewer honest nodes than the quorum size.
In this case, malicious nodes can collectively remain silent to prevent valid blocks from collecting quorum size of votes, thus obstructing consensus.
To address this issue, we aim to \emph{ensure that honest nodes are more than or equal to the quorum size} with a high probability. 
We accomplish this through rigorous theoretical derivations and parameterization, as detailed in Section~\ref{sec:epoch_security}.
As a result, even if malicious nodes are silent, they cannot prevent a valid proposer block from getting enough votes from honest nodes and passing consensus.

\subsection{Overhead Reduction}
\label{subsec:overhead}
In this section, we focus on reducing overhead while ensuring system security. 
To prevent equivocation and transaction manipulation attacks, the strawman design in Section~\ref{sec:layered_consensus} necessitates each FC to maintain states of multiple PSs to verify raw transactions, resulting in considerable overhead. 
To address the overhead issue, we first aim to \emph{leverage intra-shard consensus in each PS to ensure transaction validity}, thus preventing manipulation attacks.
To achieve this, we configure the size of each shard based on rigorous theoretical calculations (Section~\ref{sec:epoch_security}) to ensure that $quorum \ size >\# \ of \ malicious\ nodes$ with extremely high probability. 
In this case, blocks containing invalid transactions cannot collect enough votes to pass consensus, as the honest nodes will not vote for them.
However, a corrupted PS can still attack safety by equivocation (i.e., forking).
We then \emph{leverage FCs to resolve forking} issues (detailed in Section \ref{sec:layered_consensus}), detectable by examining the block header. 

\vspace{6pt}
\noindent
\textbf{Optimal Quorum Size.}
Combining the inequality that $\# \ of \ honest \ nodes$ $ \geq quorum \ size $ which prevents silent attack in Section~\ref{subsec:stagnation} and the inequality that $quorum \ size >\# \ of \ malicious\ nodes$ which prevents manipulation attack above, we derive the \textbf{\emph{optimal quorum size $m/2 + 1$ for a PS of size $m$.}}
Consequently, we carefully adjust PS sizes to ensure the proportion of malicious nodes is less than $1/2$ with a high probability according to Section~\ref{sec:epoch_security}. 
This allows honest nodes in a PS to \emph{broadcast only the block header} they voted for to the FC, who then checks signatures within the received block header and guarantees the finalized proposer block will not form forks.

\begin{figure}[t]
\centerline{\includegraphics[width=0.48\textwidth]{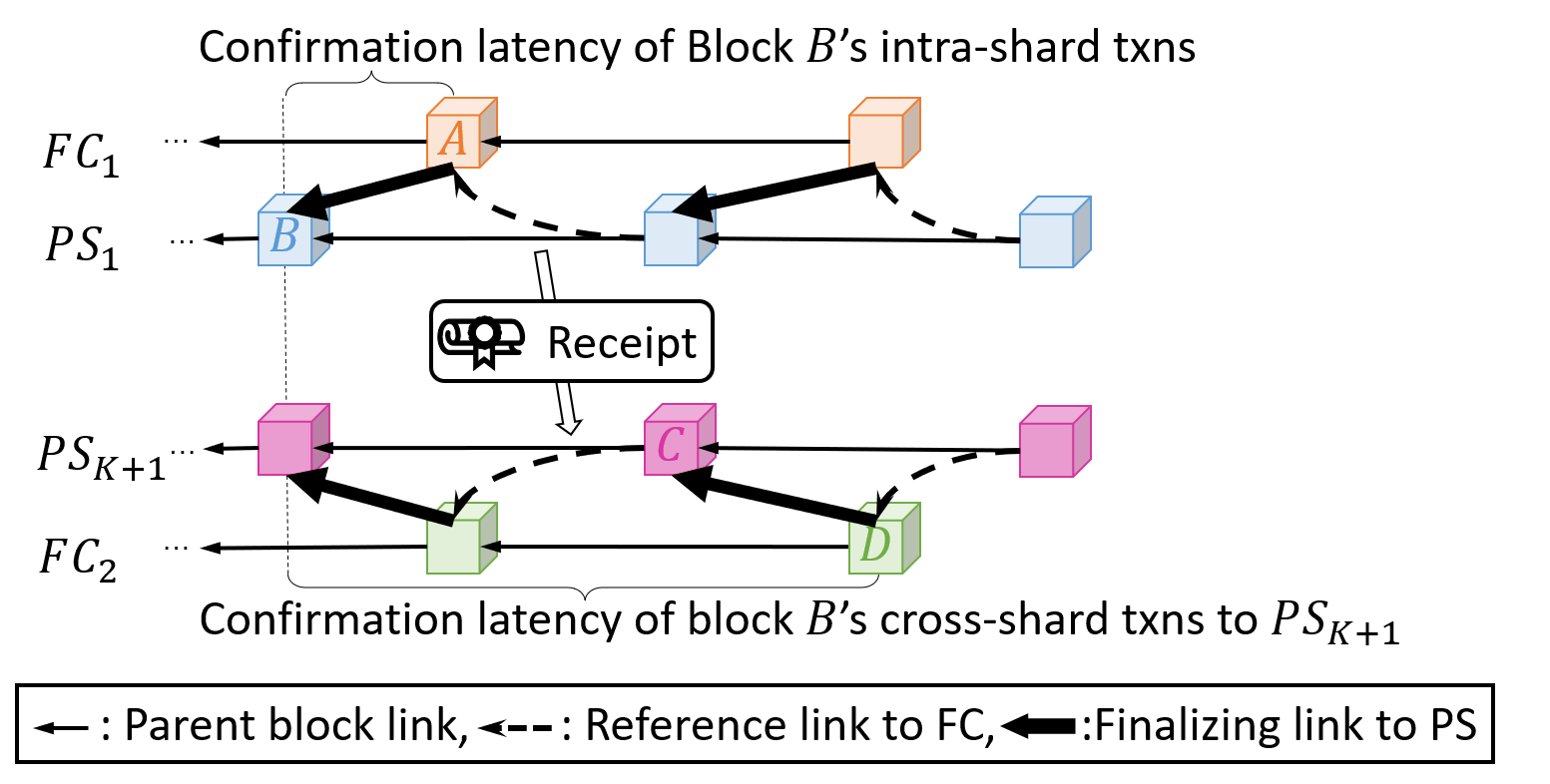}}
\caption{Process of cross-shard transactions confirmation.}
\label{fig:cross-shard}
\end{figure} 

\subsection{Cross-shard Transaction Handling}
\label{subsec:cross-shardconfir}
In blockchain systems, managing cross-shard transactions is crucial. 
Our system encounters a distinct challenge from potentially corrupted PSs: the possibility of PS blocks discarding, leading to the reversal of payment operations for cross-shard transactions originating from the source shard.
To tackle this challenge, our approach mandates that PSs await finalization by their respective FCs before proceeding with cross-shard transaction handling. 
Thanks to the security of FCs, finalized PS blocks safeguard the secure execution of cross-shard transaction deductions.

In particular, a cross-shard transaction is initially packaged into a block by the payer's PS (i.e., source PS) to execute the deduction operation. 
Upon the block production by the source PS through consensus, it still awaits finalization by the corresponding FC (i.e., source FC). 
Subsequently, nodes within the source PS are required to transmit \emph{receipt} to the destination shard, which includes the original cross-shard transaction, accompanied by proof of its existence within the source PS block and the presence of this PS block within the source FC block.
The destination shard verifies the receipt to ensure that the cross-shard transaction has been securely deducted and packages the transaction for depositing.
Finally, the cross-shard transaction is confirmed once contingent upon the finalization of the destination PS block by the destination FC.

We illustrate an example in Figure~\ref{fig:cross-shard} that confirms the cross-shard transactions (transferred from accounts in $PS_1$ to accounts in $PS_{K+1}$) included in block $B$.
Initially, $PS_1$ conducts intra-shard consensus on block $B$, including transactions whose payees are in $PS_{K+1}$, and waiting for the finalization of block $B$.
Then, the honest voter of block $B$ must broadcast a receipt to convince $PS_{K+1}$ that the deduction operation is confirmed (finalized).
The receipt includes a Merkle proof (refer to \cite{merkleTree} for details) to ensure the existence of the batch of cross-shard transactions in block $B$, and the header of block $A$ to ensure block $B$ is finalized.
Then, the destination $PS_{K+1}$ records the batched cross-shard transactions in block $C$ after verifying the receipt.
Finally, the cross-shard transactions are confirmed when block $C$ is finalized by block $D$.

\vspace{6pt}
\noindent
\textbf{Discussion.}
Our mechanism is inspired by existing relay-based cross-shard transaction processing schema (see Section~\ref{subsec:blockchain_sharding}), which has been proven for security and eventual atomicity. 
The difference lies in using secure FCs to safeguard the confirmation of cross-shard transactions.

In our system, three points guarantee the security of cross-shard transactions.
Firstly, the confirmation of the deduction operation is secure since we guarantee the security of FCs after rigorous theoretical derivations and proof, as shown in Section~\ref{sec:epoch_security}.
Secondly, using the PS and FC identification for participants (derived from epoch randomness and nodes' identities recorded in the beacon chain), nodes can verify whether blocks from other PSs and FCs have passed the consensus.
Thirdly, malicious nodes can not manipulate the batched cross-shard transactions since they cannot forge a receipt's Merkle proof.
%

Besides, the following two points guarantee the eventual atomicity of cross-shard transactions.
(I), at least one honest voter broadcasts each finalized proposer block to the destination PS since we have ensured that $quorum \ size >\# \ of \ malicious\ nodes$.
(II), the cross-layer view change replaces malicious leaders, so a well-behaved leader will eventually pick transactions in receipts.

\subsection{Pipelining Mechanism}
\label{subsec:pipeline}
In StableShard, the efficiency of the DoLC protocol is crucial. 
A naive approach would be having PSs wait for their respective FC's finalizer block before proposing new blocks, and vice versa, which leads to mutual blocking between layers. 
We employ a pipelined strategy to ensure efficient DoLC protocol operation.
This strategy enables PSs to propose blocks optimistically without waiting for finalization, while FCs consecutively finalize multiple proposer blocks from each PS without delay.
Firstly, we enable PSs to \emph{conduct block proposal optimistically}. 
PS nodes can vote for multiple proposer blocks without waiting for finalization, temporarily maintaining states post-execution. 
Secondly,  we make FCs to \emph{conduct block finalization without waiting}. 
FCs can consecutively produce finalizer blocks, finalizing multiple proposer blocks of varying heights from each of their PSs. 
However, FCs should guarantee that all chained proposer blocks have passed intra-shard consensus.
In this case, the accumulated proposer blocks in one PS can be finalized simultaneously.


 \begin{figure}[t]
\centerline{\includegraphics[width=0.48\textwidth]{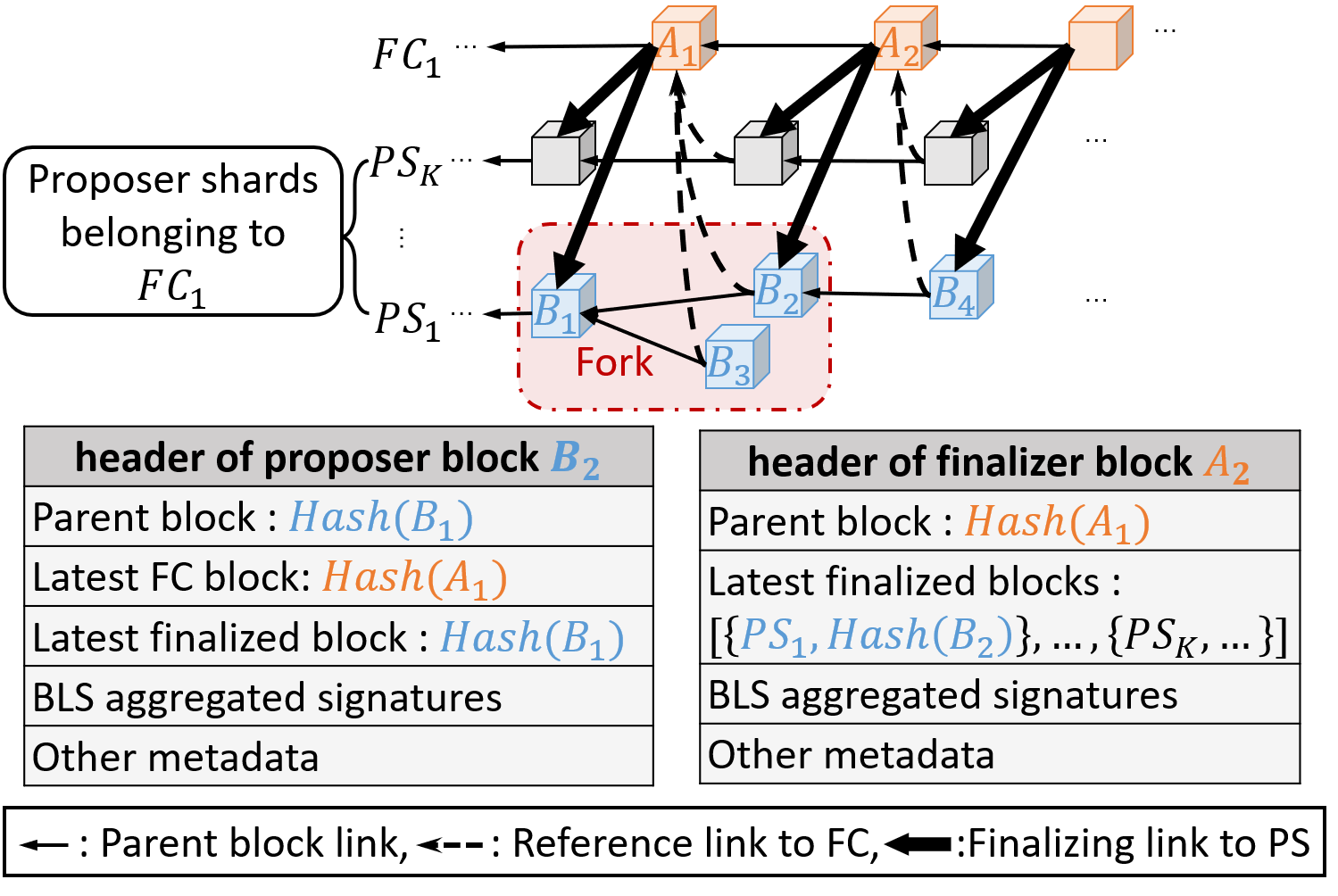}}
\caption{DoLC protocol and block header design.}
\label{fig:layered_consensus}
\end{figure} 

\subsection{Summary of StableShard}
This section summarizes our system using an example shown in Figure~\ref{fig:layered_consensus}. 
We assume that $K$ PSs from $PS_1$ to $PS_K$ belong to $FC_1$, focusing on the finalization process of $FC_1$ to $PS_1$.
The optimal quorum size for each PS of size $m$ is $m/2+1$, with a toleration threshold of less than $m/2$. 
The \emph{block proposal} phase takes place inside $PS_1$ continuously since we guarantee the liveness of PSs.
$PS_1$ can only create valid proposer blocks via intra-shard consensus so that the \emph{cross-layer verification} only involves block headers.
However, blocks $B_2$ and $B_3$ are valid but form a fork after block $B_1$.
$FC_1$ receives the block headers broadcast by honest nodes in $PS_1$ and selects only one fork branch (e.g., block $B_2$ was received first, hence chosen for finalization) to finalize.
Sequentially, $PS_1$ can only link the new block to $B_2$ after receiving finalizer block $A_2$.
As a result, $PS_1$ safely confirms the intra-shard transaction and transmits receipts (containing the headers of block $B_2$ and $A_2$) of cross-shard transactions in block $B_2$.

\section{Security Analysis and Discussion}
\label{sec:analysis}

In this section, we begin by computing the failure probability of StableShard within each epoch. 
With negligible failure probability (i.e., ensuring the epoch security with extremely high probability), we then prove that the system is guaranteed to be secure as long as the proportion of malicious nodes in each PS is less than $1/2$ and that in each FC is less than $1/3$.

\subsection{Epoch Security}
\label{sec:epoch_security}
To guarantee epoch security, we must ensure the system failure probability is below a certain threshold.
This section aims to derive the failure probability of \mbox{StableShard} in each epoch so that we can adjust the system parameters to ensure a negligible system failure probability.
We require that the fractions of malicious nodes are less than $1/3$ within each finalizer committee and less than $1/2$ within each proposer shard during each epoch.
The overview of the calculation is as follows.
Firstly, we calculate the failure probability of a finalizer committee. 
Specifically, it is divided into two cases.
\textbf{Case 1}: It contains at least $1/3$ fraction of malicious nodes.
\textbf{Case 2}: It is honest, but its proposer shards contain at least $1/2$ fraction of malicious nodes.
Secondly, as in existing work \cite{rapidchain, cochain, li2022jenga, gearbox}, we calculate the union bound over all FCs to bound the failure probability of the whole system.

We calculate the failure probability for a finalizer committee now.
The network size, the fraction of malicious in the network, and the size of finalizer committees are denoted as $N,f,n$, respectively.
Let $X$ denote the random variable of the number of malicious nodes in a finalizer committee.
We leverage the hypergeometric distribution, similar to existing sharding blockchains~\cite{sgxSharding, pyramid, rapidchain}, to calculate the probability of a finalizer committee has $X = x$ malicious nodes:
\begin{equation}
Pr[X = x]
\\
= \frac{
\binom{f\cdot N}{x} \binom{N-f\cdot N}{n-x}}{\binom{N}{n}
}.
\label{equ_fc_has_x_malicious}
\end{equation}

Based on expression~\ref{equ_fc_has_x_malicious}, when $x\geq n/3$, the finalizer committee fails according to \textbf{\emph{case 1}}.
The probability is:
\begin{equation}
Pr[X \geq n/3]
\\
=\sum_{x= \lfloor n / 3\rfloor }^{n} Pr[X = x].
\label{equ_FC_fail_case1}
\end{equation}

Let $Y$ denote the random variable of the number of malicious nodes in the proposer shard of size $m$.
The probability of a PS with malicious nodes not less than $1/2$ fraction in an honest FC is:
\begin{equation} 
Pr[ Y \geq m/2 | X < n/3]
\\
=\sum_{x= 1 }^{\lfloor n / 3 \rfloor - 1} \sum_{  y= \lfloor m / 2\rfloor  }^{m} 
\frac{\binom{x}{y} \binom{n-x}{m-y}}{\binom{n}{m}}
.
\label{equ_ps_fail_when_fc_X} 
\end{equation}

Like prior research~\cite{rapidchain, pyramid, cochain}, we also calculate the upper bound failure probability, assuming each shard's failure probability is independent.
We calculate the union bound over $K=n/m$ PSs, which results in 
the upper bound of the failure probability of an honest FC (according to \textbf{\emph{case 2}} that exists at least one PS containing at least $1/2$ malicious nodes).
\begin{equation} 
Pr[ \exists Y \geq m/2 | X < n/3]
\\
\leq  K \cdot Pr[  Y \geq m/2 | X < n/3].
\label{equ_FC_fail_case2}
\end{equation} 

Combine expressions \ref{equ_FC_fail_case1} and \ref{equ_FC_fail_case2}, we get the upper bound of the failure probability of a FC:
\begin{equation}
Pr[ FC \ Failure]
\\
\leq Pr[X \geq n/3] + Pr[ \exists Y \geq m/2 | X < n/3]
\label{equ_fc_fail}
\end{equation}

We calculate the union bound over $C=N/n$ FCs to bound the failure probability of the system, similar to the calculation of the upper bound of a FC's failure probability.
\begin{equation}
Pr[ System \ Failure]
\\
\leq C \cdot Pr[ FC \ Failure].
\label{equ_sys_fail}
\end{equation}

\noindent
\textbf{Ensuring Negligible Failure Probability.} 
StableShard, like most previous blockchain sharding works~\cite{omniledger, rapidchain, sgxSharding, pyramid}, must ensure a negligible failure probability within each epoch to maintain security. 
Based on expressions~\ref{equ_FC_fail_case1} and \ref{equ_ps_fail_when_fc_X}, we need to adjust the FC size $n$ and the PS size $m$ to make sure there is a small $\varepsilon$ existing so that:
\begin{equation}
C \cdot ( Pr[X \geq n/3] +  K \cdot Pr[  Y \geq m/2 | X < n/3])
\leq \varepsilon.
\label{final_fail_pro}
\end{equation}
If we achieve a negligible $\varepsilon$ such that the expression~\ref{final_fail_pro} holds, epoch security is ensured.
The specific settings and the probabilities are shown in Section~\ref{sec:para_setting}.

\subsection{Security Analysis of DoLC Protocol}
\label{protocol_security}
In this section, we define and prove the security of our DoLC protocol, encompassing both safety and liveness aspects.


Since each node is assigned into one proposer shard and one finalizer committee simultaneously, let $\mathrm{C}_t^{P,\mathrm{n}}$ and $\mathrm{C}_t^{F,\mathrm{n}}$ denote the proposer chain and finalizer chain output by a full node $\mathrm{n}$ at time $t$.  
If the blocks, recorded from the genesis block onward, that constitute chain $\mathrm{C}_t^{\mathrm{n}}$ are a subset of the blocks that constitute chain $\mathrm{C}_{t'}^{\mathrm{n'}}$, it is denoted as $\mathrm{C}_t^{\mathrm{n}} \preceq \mathrm{C}_{t'}^{\mathrm{n'}}$.
Additionally, if the blocks constituting chain $\mathrm{C}_t^{\mathrm{n}}$ are a proper subset of the blocks constituting chain $\mathrm{C}_{t'}^{\mathrm{n'}}$, it is denoted as $\mathrm{C}_t^{\mathrm{n}} \prec \mathrm{C}_{t'}^{\mathrm{n'}}$.
The DoLC protocol is secure if the following two properties are satisfied.
\begin{itemize}
  \item \textbf{Safety:} Safety requires that the finalized blocks will not fork and only contain valid transactions. 
  For any times $t, t'$, and any two honest nodes $\mathrm{n},\mathrm{n'}$ from the same proposer shard, either $\mathrm{C}_t^{P,\mathrm{n}} \preceq \mathrm{C}_{t'}^{P,\mathrm{n'}}$ or vice versa; 
 
  \item \textbf{Liveness:} Liveness requires the system to finalize blocks continuously.
  For any time $t$ and any honest node $\mathrm{n}$, there exists a finite delay $\Delta$ such that $\mathrm{C}_t^{P,\mathrm{n}} \prec \mathrm{C}_{t+\Delta}^{P,\mathrm{n}}$.
\end{itemize}

\begin{theorem}
\label{Theorem3} 
In DoLC protocol, considering a finalizer committee $FC_i$ ($1 \leq i \leq C$) and one of its corresponding proposer shard $PS_j$ ($(i-1)K+1\leq j \leq iK$), which has $m_j$ nodes, and a fraction $f_j$ being malicious. 
assuming the network is a partial-synchronous network, the DoLC protocol instantiated with finalizer committee $FC_i$ and proposer shard $PS_j$ satisfies

\begin{itemize}
  \item safety iff $FC_i$ is safe and each of its constituent proposer shard $PS_j$ is running FBFT with $f_j < 1/2 $ and a quorum of $q_j = m_j/2 +1$;
  \item liveness iff $FC_i$ is live and each of its constituent proposer shard $PS_j$ is running FBFT with $f_j < 1/2$ and a quorum of $q_j = m_j/2 +1$.
\end{itemize}
\end{theorem}

\begin{proof}
We first prove the safety.
Specifically, we first prove that finalized blocks are free from forks. Subsequently, we demonstrate the validity of transactions within the finalized blocks.
Suppose the finalizer committee is safe. 
Then, without loss of generality, $\mathrm{C}_{t_1}^{F, \mathrm{n_1}} \preceq \mathrm{C}_{t_2}^{F, \mathrm{n_2}}$ for any two nodes $\mathrm{n_1}$ and $\mathrm{n_2}$ from this finalizer committee and times $t_1$ and $t_2$.
Let $B_t^P$ represent the latest block of $PS_j$ finalized by $FC_i$ at time $t$.
As detailed in Section~\ref{sec:layered_consensus}, honest nodes within $FC_i$ play a crucial role in cross-layer verification. 
They ensure that, during the consensus process, the most recent finalized proposer block (e.g., $B_{t'}^P$) remains free from conflicts with the blocks forming the chain that concludes with $B_t^P$ (i.e., $B_t^P \preceq B_{t'}^P$ and $t < t'$). 
And since blocks are linked by the collision-resistant hash function, the sequence of the finalized proposer block observed by $\mathrm{n_1}$ is a prefix of $\mathrm{n_2}$'s sequence.
It implies that  $\mathrm{C}_t^{P,\mathrm{n_1}} \preceq \mathrm{C}_{t'}^{P,\mathrm{n_2}}$, concluding the safety proof.
Besides, when the proposer shard is running FBFT~\cite{Harmony} with $m_j$ distinct nodes and a quorum of $q_j = m_j/2 + 1$, each proposer block is verified by at least one honest node, thus even if the finalizer committee only verify the header of proposer block, the validity of the raw transaction is preserved. 

When we prove the liveness, we first ensure liveness when malicious nodes stay silent and then confirm liveness when a malicious leader avoids proposing valid blocks. 
In a proposer shard $PS_j$ with $f_j< 1/2$ malicious nodes, there are at least $m_j/2 + 1$ honest nodes.
A quorum size of $m_j/2 + 1$ within $PS_j$ can be reached from honest nodes within a bounded delay, even if all malicious nodes remain silent under a partial-synchronized network.
Assuming the finalizer committee remains live.
Then, without loss of generality, there exists a finite delay $\eta$ such that $\mathrm{C}_{t}^{F, \mathrm{n}} \prec \mathrm{C}_{t+\eta}^{F, \mathrm{n}}$ for any node $\mathrm{n}$ from this finalizer committee at any time $t$.
Consequently, any proposer block produced will eventually be finalized. 
Therefore, liveness is guaranteed when an honest leader proposes valid blocks.
Suppose the latest finalized proposer block is $B_t^P$, and $PS_j$'s malicious leader proposes $B_{t'}^P$, $B_t^P \not\prec B_{t'}^P$.
In this case, $FC_i$ can not finalize $B_{t'}^P$.
Even if malicious nodes in $PS_j$ remain silent, the $FC_i$ will collect quorum size of $m_j/2+1$ \emph{Complaint} messages from honest nodes within finite delay under a partial-synchronous network.
Due to the finite time in which $FC_i$ reaches consensus upon the leader replacement request from $PS_j$, and given the limited number of nodes (leader candidates) within the $PS_j$, there will be an honest node being selected as a leader within finite delay $\mu$ to propose a new block $B_{t+\mu}^P$, $B_t^P \prec B_{t+\mu}^P$. 
If $PC_j$ keeps proposing valid blocks as the descendant of the latest finalized block, the finalizer committee will finalize such valid proposer block within a bounded delay to expand the proposer chain, guaranteeing liveness.
\end{proof} 
In the Theorem~\ref{Theorem3}, we assume that FC is secure, and now we illustrate the specific requirement that guarantees FC is secure, when FC adopts FBFT as its consensus protocol.
\begin{proposition}
\label{propostition}
FBFT~\cite{Harmony} has the same security guarantee with  PBFT~\cite{pbft}, which is stated as follows.
PBFT satisfies safety and liveness for a group of $g$ nodes using $2g/3 + 1$ as a quorum when the fraction of byzantine nodes is less than $1/3$.
\end{proposition}
Combining Theorem~\ref{Theorem3} and Proposition~\ref{propostition}, we now give a complete corollary for safety and liveness.
\begin{corollary}
The DoLC protocol, instantiated with finalizer committee $FC_i$ running FBFT with $g_i$ nodes ($FC_i$ contains $f_i$ fraction of malicious nodes and uses $q_i = 2g_i/3+1$ as a quorum), and proposer shard $PS_j$ running FBFT~ with $m_j$ nodes ($PS_j$ contains $f_j$ fraction of malicious nodes and uses $q_j=m_j/2+1$ as a quorum), under partial-synchronous network, satisfies
\begin{itemize}
  \item safety iff $f_i<1/3$ and $f_j<1/2$;
  \item liveness iff $f_i<1/3$ and $f_j<1/2$.
\end{itemize}
\end{corollary}

\subsection{Discussions}
\label{subsec:additional_discussion}

\noindent
\textbf{Temporary Stagnation.}
StableShard can prevent temporary stagnation, which arises from reshuffling and ledger migration of corrupted shards.
The rationale behind this is that we guarantee the liveness of corrupted shards.
Nevertheless, StableShard cannot prevent the delay of the rotation of a malicious leader.
Fortunately, it only incurs a minor delay, as it does not involve cross-shard ledger migration.
Besides, several studies have been conducted to mitigate the impact of malicious leaders on sharding, and these solutions can be applied to our system. 
For instance, RepChain~\cite{repchain} introduces a reputation mechanism to effectively reduce the possibility of a node with malicious behavior being elected as a leader.

\vspace{6pt}
\noindent
\textbf{Additional Attacks.}
StableShard can also resist some other type of attacks.
For instance, the eclipse attack can isolate the leader of PS and FC by controlling all their incoming and outgoing connections.
Fortunately, our cross-layer view change mechanism can replace a PS leader, and the original FBFT comes with a view change mechanism to replace the FC leader.
In addition, malicious nodes may collude to replace the honest leader.
For two reasons, malicious nodes cannot replace a well-behaved, honest leader.
Firstly, malicious nodes cannot obtain the honest leader's private key to forge evidence of attacks.
Additionally, they cannot gather quorum size of \texttt{Complain} messages because other honest nodes will not follow suit.
The attackers can launch transaction censorship attacks to censor transactions~\cite{sandwich} intentionally.
Red Belly~\cite{redBelly} proposes a leaderless BFT-typed consensus protocol.
It prevents the transaction censorship attack conducted by a single leader via merging the micro-blocks proposed by multiple nodes into one complete block.
Moreover, it can also prevent the eclipse attack on the specific leader.
Most importantly, we can also use the above consensus protocol to resist these attacks since the intra-shard consensus protocol is an alternative component.

\vspace{6pt}
\noindent
\textbf{Cross-shard Transaction.}
Besides supporting transfer transactions, StableShard can also exploit exiting methods \cite{monoxide, pyramid, Harmony} to handle 
complex smart contracts that span multiple proposer shards (which can also be in different FCs).
The rationale is that StableShard provides safe finalization of proposer blocks to confirm each execution step, similar to the finalization of deduction operation for cross-shard transfer transactions (refer to Section~\ref{subsec:cross-shardconfir}).
The optimization of complex smart contracts processing is orthogonal to our study and can be considered our future work.

Although increasing the number of shards may result in more cross-shard communications, the enhanced concurrency achieved through smaller shards outweighs the potential performance impact. 
The main reason is that the size of a cross-shard transaction (typically receipts or metadata) is usually smaller than an intra-shard transaction \cite{Harmony, monoxide}.
Moreover, StableShard only transmits headers for cross-layer verification.
\section{Implementation and Evaluation}

\subsection{Implementation}

We developed a StableShard prototype in Golang based on Harmony\cite{Harmony}, 
a prominent industrial permissionless blockchain sharding project that was once among the top 50 cryptocurrencies by market cap.
The \ac{FBFT} consensus protocol, proposed by Harmony, is used as our intra-shard consensus to ensure a fair comparison to demonstrate \mbox{StableShard's} performance improvements.
%
Harmony, a traditional blockchain sharding unable to tolerate corrupted shards, serves as our baseline protocol. 
We also implemented a GearBox (CCS 22) \cite{gearbox} prototype as a state-of-the-art comparison.
GearBox initially sets shards to a small size, prone to losing liveness. 
Upon detecting a corrupted shard, it reorganizes it until liveness is restored.

\begin{figure*}[t]
\centering{}
\subfloat[Total TPS comparison]{\includegraphics[scale=0.3]{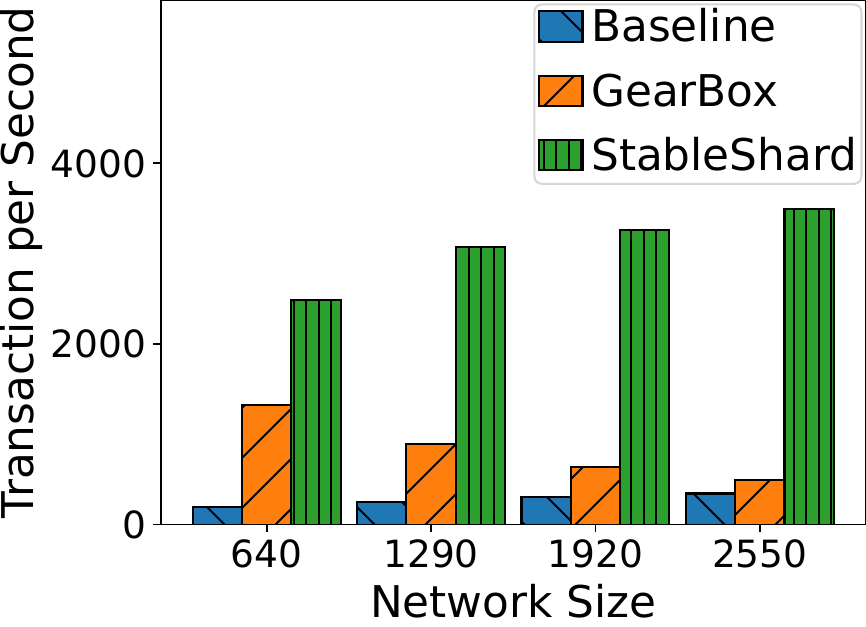}\label{fig:overallTPS}}
\hfill
\subfloat[Per-shard TPS comparison]{\includegraphics[scale=0.3]{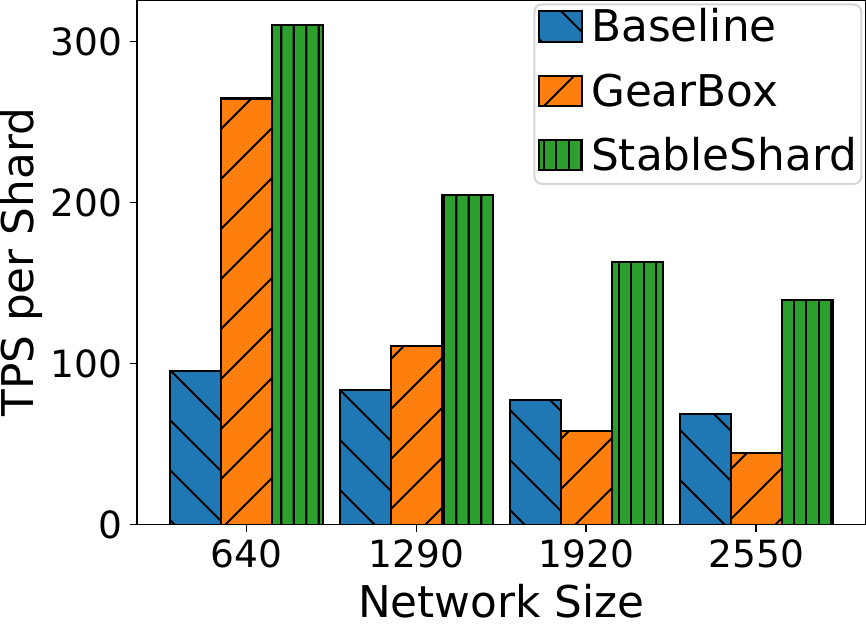}\label{fig:perShardTPS}}
\hfill
\subfloat[Confirmation latency comparison]{\includegraphics[scale=0.3]{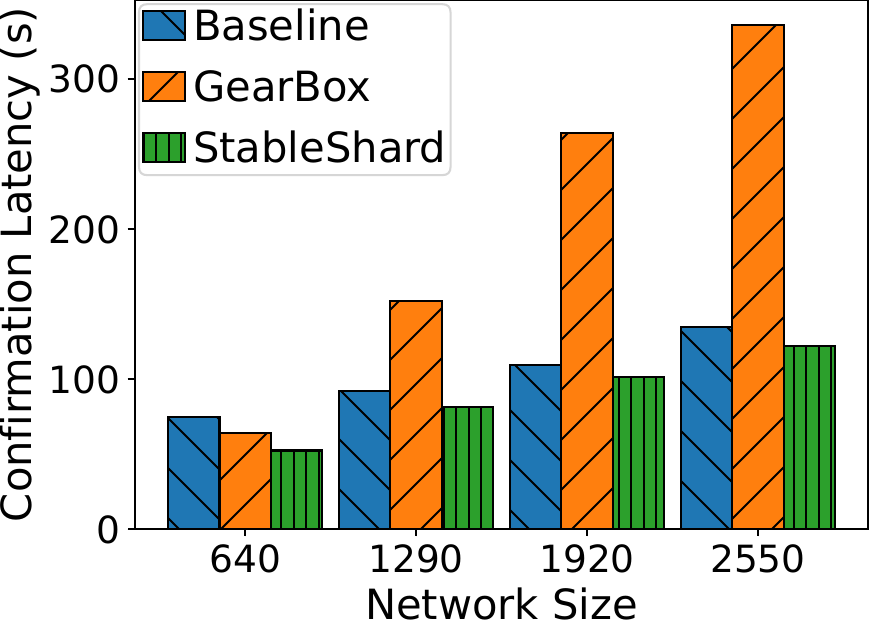}\label{fig:overallLatency}} 
\hfill
\subfloat[Latency breakdown]{\includegraphics[scale=0.3]{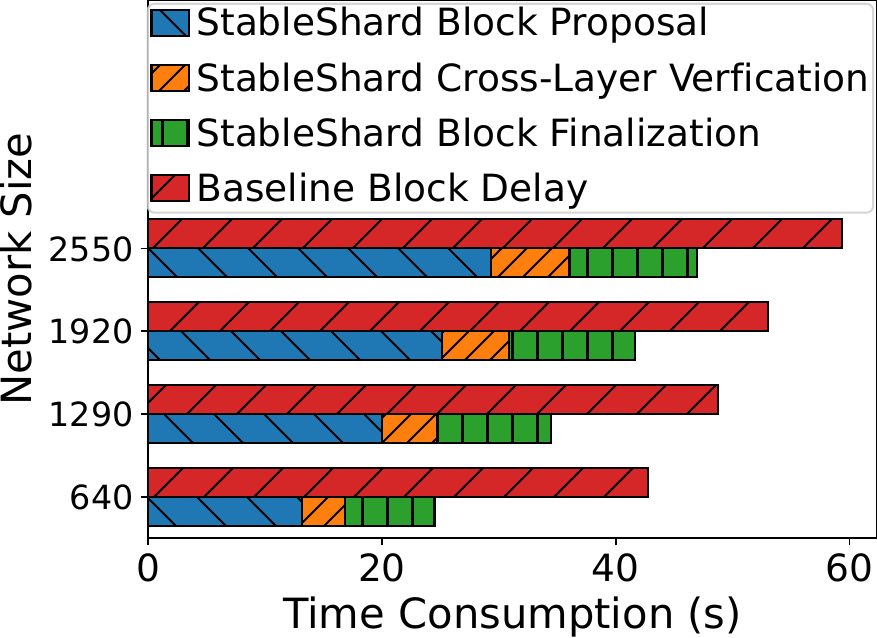}\label{breakdown1}} 
\centering
\caption{Performance under various network scales.}
\end{figure*}


\subsection{Experimental Setup}
We use a large-scale network of 2,550 nodes on 12 Amazon EC2 instances for testing. 
Each instance has a 128-core CPU and 50 Gbps network bandwidth, hosting up to 213 nodes. 
Docker containers and Linux Traffic Control managed inter-node communication, enforcing a 100 ms message delay and a 50 Mbps bandwidth limit per node. 
The experiment uses 512-byte transactions, and each block accommodates up to 4,096 transactions (i.e., 2MB blocks) as in existing work~\cite{rapidchain}. 
The transactions are based on historical data of Ethereum \cite{xblock}, in which the proportion of cross-shard transactions increases with the number of shards.
The total fraction of malicious nodes is set at $f=1/4$, typical in practical network environments~\cite{buildingBlocks_survey}.

\subsection{Parameter Settings}
\label{sec:para_setting}
In StableShard, proposer shard and finalizer committee sizes should be adjusted to guarantee a negligible system failure probability, as mentioned in Section~\ref{sec:epoch_security}.
In the baseline (Harmony), we leverage the equations based on classical hypergeometric distribution in~\cite{rapidchain} to determine shard sizes. 
We leverage the function proposed by GearBox and conduct 10 million simulations from the beginning of the epoch to get the average number of shards for implementing GearBox.
The negligible failure probability is less than $2^{-17}\approx7.6\times10^{-6}$, the same as many existing works~\cite{pyramid,li2022jenga}, meaning that one failure occurs in about 359 years (with one-day epoch).

The parameter settings are shown in Table~\ref{table:parameter}.
To guarantee the security and scalability of finalizer committees, we set the size of FCs the same as the shard size in the baseline.
After that, we set the proposer shard size to the minimum value for better performance.
We recommend that developers reduce FC and PS size for better performance without compromising security, according to Section~\ref{sec:epoch_security}.
According to Table~\ref{table:parameter}, StableShard significantly reduces the PS size and increases the number of PS compared with the baseline.
While PS sizes in StableShard are marginally larger than GearBox's shard sizes, StableShard can accommodate more PSs. 
This is because the probability of system failure rises with the number of shard samples. 
In StableShard, the number of PSs is equivalent to the number of samples since we do not reshuffle corrupted shards. 
However, the number of shards in GearBox is lower than the number of samples, as GearBox requires multiple re-sampling attempts to obtain a single shard.

\begin{table}[tbh]
\caption{Parameter settings.}
\centering{}%
\begin{tabular}{c|c|c|c|c}
\hline
\hline
Network Size & 640 & 1,290 & 1,920 & 2,550 \tabularnewline
\hline 
\hline
\multicolumn{5}{c}{Baseline (Harmony)}\tabularnewline
\hline 
\# of Shards  & 2& 3 & 4 & 5 
\tabularnewline
\hline 
 Shard Size  &320 & 430 &  480 &  510 
\tabularnewline
\hline 
Failure Probability ($\cdot10^{-6}$) & 2.8 & 3.3 & 4.8& 4.6 \tabularnewline 
\hline
\hline
\multicolumn{5}{c}{GearBox}\tabularnewline
\hline 
\# of Shards & 5 & 8 & 11 & 11\tabularnewline
\hline 
Avg. Shard Size &  65 &  75 & 76 &  80 \tabularnewline
\hline 
Failure Probability ($\cdot10^{-6}$) & 3.4 & 3.89 & 4.1 & 3.4 \tabularnewline 
\hline
\hline
\multicolumn{5}{c}{StableShard}\tabularnewline 
\hline 
\# of FCs  & 2 & 3 & 4 & 5 \tabularnewline
\hline 
 FC Size  & 320 &  430 & 480 &  510 \tabularnewline
\hline 
\# of PSs per FC  & 4&5   &5& 5 \tabularnewline
\hline 
 PS Size  &  80&  86   &  96&  102 \tabularnewline
\hline
total \# of PSs & 8 & 15 & 20 & 25 \tabularnewline
\hline 
Failure Probability($\cdot10^{-6}$)   & 4.3 & 6.0 & 5.8 & 6.8\tabularnewline 
\hline
\hline
\end{tabular}
\label{table:parameter}
\end{table}

\subsection{Transaction Throughput and Latency}
\label{sec:evaluation_overall_performance}
We first compare the throughput (transaction per second, TPS) of StableShard and other systems at different network sizes, shown in Figures~\ref{fig:overallTPS} and~\ref{fig:perShardTPS}.
Compared with Harmony, StableShard has more PSs with smaller size, thus achieving more than 10 times the total TPS and 2 times the TPS for a single shard.
Compared with GearBox, StableShard has more PSs and avoids network-wide consensus via scalable DoLC protocol, thus achieving up to 7 times the total TPS and 3 times the TPS for a single shard in a network size of 2,550.

 
We evaluate average transaction confirmation latency (i.e., the duration between a transaction starts to be processed until it is finalized, similar to previous works~\cite{pyramid}), as shown in Figure~\ref{fig:overallLatency}.
Unlike the considerable gains in throughput, the transaction confirmation latency of StableShard is only a few seconds lower than that of Harmony.
The reasons are that StableShard has a larger proportion of cross-shard transactions, and the confirmation of transactions awaits FCs' finalization.
However, StableShard reduces latency significantly compared with GearBox due to our DoLC protocol, which avoids network-wide consensus and ensures scalability and efficiency. 

\subsection{Breakdown of Block Latency}
We analyze the block latency in StableShard's three-phase DoLC protocol:  block proposal in PSs, cross-layer verification, and block finalization in FCs. 
Unlike transaction confirmation latency, block latency represents the time from a proposer block's proposal to its finalization. 
As depicted in Figure~\ref{breakdown1}, StableShard's overall block latency is shorter than the baseline for three reasons. 
Firstly, fewer nodes participate in StableShard's block proposal, reducing delay. 
Secondly, the system uses block headers for efficient cross-layer verification. 
%
Finally, while the FC sizes are comparable to the baseline shards, each FC's consensus on metadata (proposer block hashes) rather than transactions accelerates block finalization.


\begin{figure}[t]
\centering{}
\subfloat[TPS]{\includegraphics[scale=0.3]{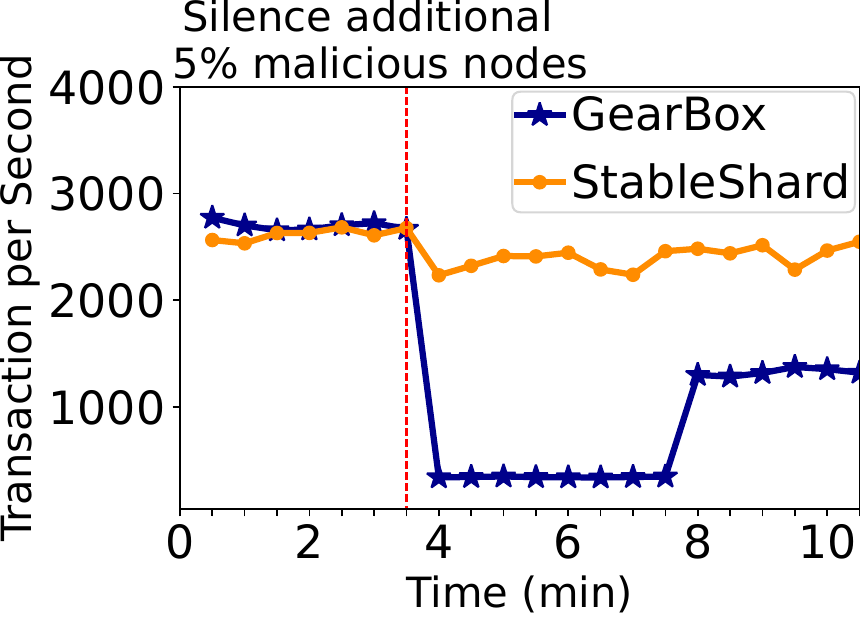}\label{fig:simluTPS}}
\hfill
\subfloat[Confirmation latency]{\includegraphics[scale=0.3]{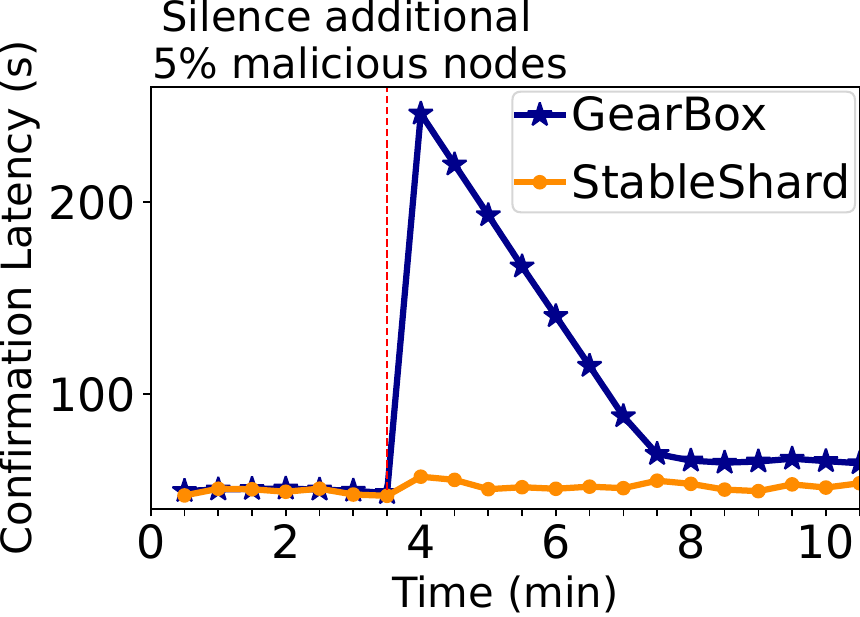}\label{fig:simluLatency}} 
\centering
\caption{Performance under silent attack.}
\end{figure}

\subsection{Temporary Stagnation under Various Malicious Fractions}
\label{subsec:eva_stag}
This experiment aims to assess StableShard's ability to provide stable performance.
We fix the network size as 640 and vary the fraction of malicious nodes to investigate its impact on system performance stability.
This situation is common in real-world systems because adversaries can arbitrarily control the fraction of their malicious nodes to hinder system processes. 
Even in the absence of malicious nodes, the instability of the blockchain network environment can also lead to different percentages of nodes becoming unresponsive and hindering the system's operation.
We assume that the adversary silences 20\% of the nodes at the beginning and silences another 5\% of the nodes to initiate silent attacks to stagnate the system at the time $T=3.5$.

\begin{figure}[t]
\centerline{\includegraphics[scale=0.3]{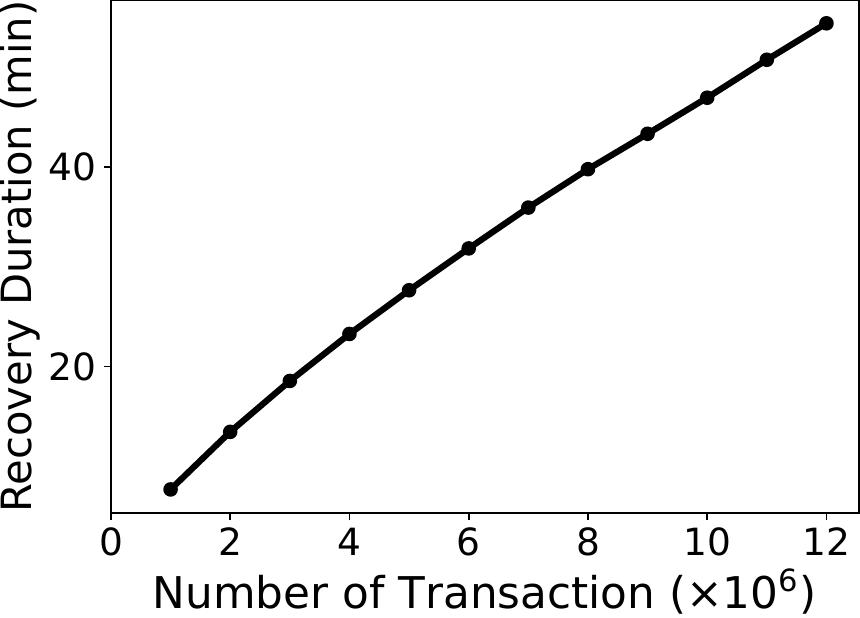}}
 \caption{Migration duration.}\label{fig:recoveryTime}
\end{figure}

As shown in Figure~\ref{fig:simluTPS}, the throughput of both systems drops at time $T$.
For StableShard, the system wait longer to reach the quorum size for consensus due to additional silent nodes after time $T$, reducing efficiency.
However, StableShard's throughput is still more than 6 times the throughput of GearBox after time $T$.
This is because most shards in GearBox lose liveness and cannot package any transactions.
GearBox then underwent nearly 4 minutes of shard re-sampling and state migration involving 1 million Ethereum transactions~\cite{xblock}.
However, the latency in real scenarios is even more significant than this and increases with state sizes, as shown in Figure~\ref{fig:recoveryTime}.
Finally, the throughput of GearBox is only partially recovered due to the increased shard size and reduced shard count.

As shown in Figure~\ref{fig:simluLatency}, the latency of both systems increases at time $T$.
For StableShard, the reason for increased latency is that it takes longer to reach a consensus within PSs or FCs when more nodes keep silent.
However, the latency of GearBox significantly increases after time $T$.
This is because all cross-shard transactions directed to the corrupted shards are stuck until the corrupted shard's liveness is recovered.

We evaluate the latency of state migration during reshuffling (only exists in GearBox) based on historical Ethereum transactions~\cite{xblock}.
As shown in Figure~\ref{fig:recoveryTime}, the more transactions executed, the longer time required for state migration.
This is due to the greater involvement of accounts, resulting in a more intricate migration state.
Note that the evaluation only involves 12 million Ethereum transactions for ten days~\cite{xblock}, and the real-world situation of Ethereum, which has been running for several years, will involve more considerable delay.
On the other hand, \emph{StableShard is immune to this time-consuming stagnation process.}



\section{Conclusion}
This paper presents StableShard, a sharded blockchain that achieves high concurrency with stable and scalable performance under corrupted shards. 
The key contribution is a PS-FC Division-of-Labor Collaboration (DoLC) protocol: many small proposer shards (PSs) maximize concurrency by proposing blocks quickly and ensuring transaction validity via quorum voting, while multiple large finalizer committees (FCs) provide scalable, header-only fork resolution and deterministic finality for multiple PSs without network-wide consensus. 
Crucially, FCs maintain stagnation-free, in-place liveness for faulty PSs through a cross-layer view-change mechanism and a jointly tuned quorum/committee configuration, thereby eliminating recovery-after-failure procedures that cause severe service interruption. 
By jointly parameterizing committee sizes and quorum thresholds, StableShard enables a Byzantine fraction approaching <1/2 in each PS while keeping each FC secure (<1/3) with overwhelmingly high probability, allowing many small PSs and multiple coexisting FCs for stable, scalable concurrency. 
A Harmony-based prototype and large-scale evaluation (up to 2,550 nodes) demonstrate up to 10x throughput improvement and significantly more stable concurrency under attacks compared with prior designs.

\bibliographystyle{abbrv}
\bibliography{Files/ref}

\begin{IEEEbiography}[{\includegraphics[width=1in,height=1.25in,clip,keepaspectratio]{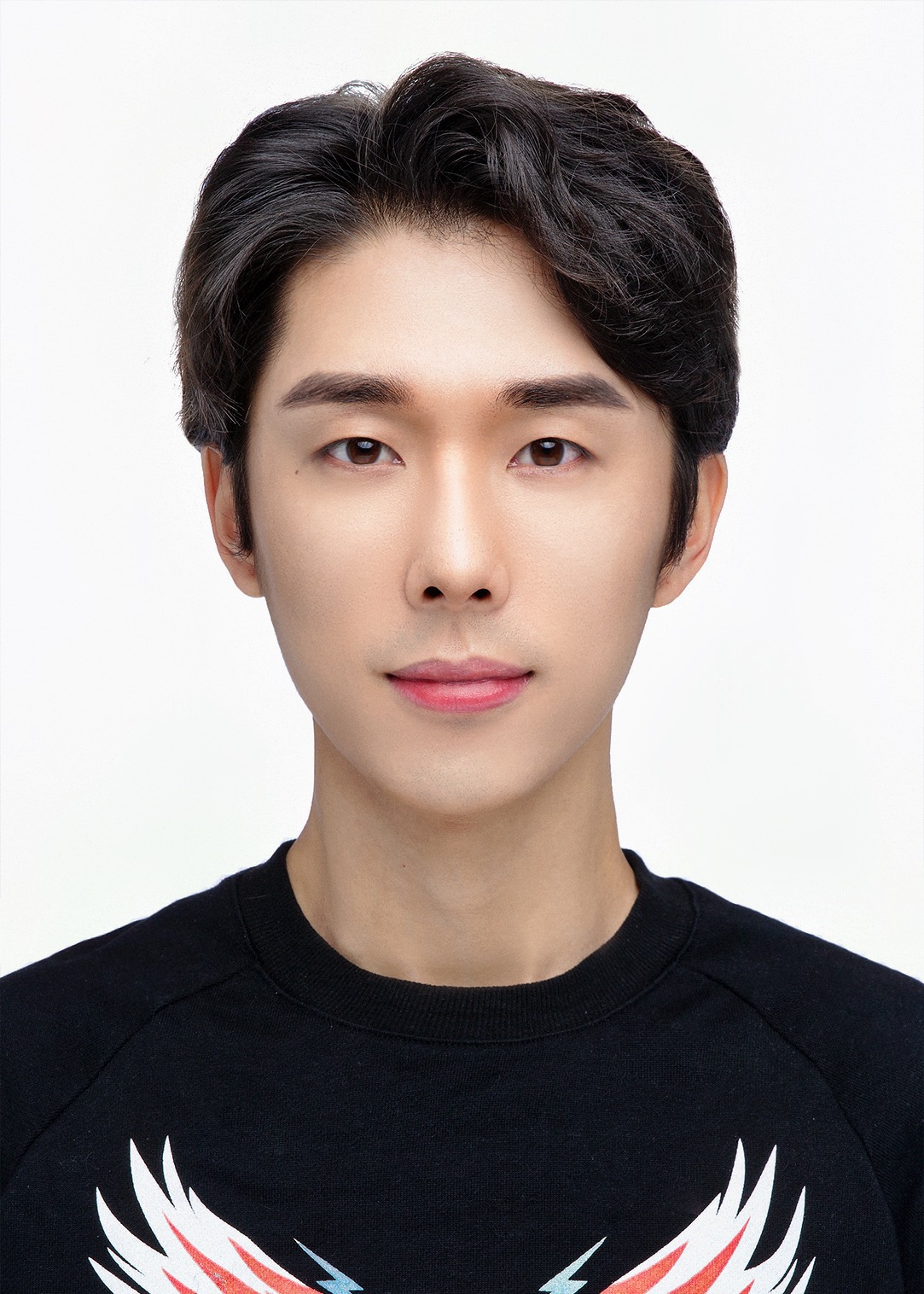}}]{Mingzhe Li}
is currently a US-equivalent Assistant Professor with the School of Computing and Information Technology, Great
Bay University.
He received his Ph.D. degree from the Department of Computer Science and Engineering, Hong Kong University of Science and Technology in 2022.
Prior to that, he received his B.E. degree from Southern University of Science and Technology.
His research interests are mainly in blockchain sharding, consensus protocol, blockchain application, network economics, and crowdsourcing.
\end{IEEEbiography}
\begin{IEEEbiography}[{\includegraphics[width=1in,height=1.25in,clip,keepaspectratio]{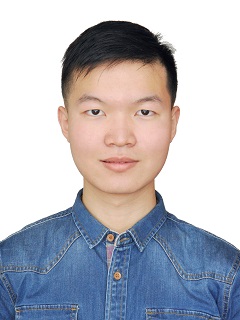}}]{You Lin}
is currently a master candidate with Department of Computer Science and Engineering, Southern University of Science and Technology. 
He received his B.E. degree in computer science and technology from Southern University of Science and Technology in 2021. 
His research interests are mainly in blockchain, network economics, and consensus protocols.
\end{IEEEbiography}
\begin{IEEEbiography}
[{\includegraphics[width=1in,height=1.25in,clip,keepaspectratio]{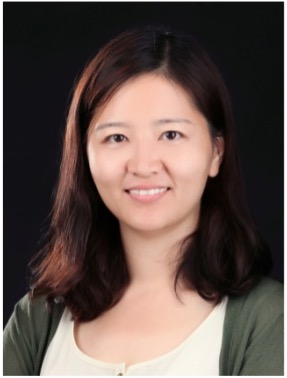}}]{Jin Zhang} 
is currently an associate professor with Department of Computer Science and Engineering, Southern University of Science and Technology. 
She received her B.E. and M.E. degrees in electronic engineering from Tsinghua University in 2004 and 2006, respectively, and received her Ph.D. degree in computer science from Hong Kong University of Science and Technology in 2009. 
Her research interests are mainly in mobile healthcare and wearable computing, wireless communication and networks, network economics, cognitive radio networks and dynamic spectrum management. 
\end{IEEEbiography}

\end{document}